\documentclass{aa}  

\usepackage{graphicx}
\usepackage{txfonts}
\usepackage{hyperref}

\hypersetup{colorlinks=true,linkcolor=blue,filecolor=blue,urlcolor=blue,citecolor=blue}
\usepackage{arydshln}
\usepackage{xcolor}
\usepackage{multirow}
\usepackage[switch]{lineno}

\makeatletter
\renewcommand*\aa@pageof{, page \thepage{} of \pageref*{LastPage}}
\makeatother

\begin{document}

   \title{Distance estimation of gamma-ray emitting BL Lac objects from imaging observations}

   \author{K. Nilsson\inst{1}\fnmsep\thanks{\email{kani@utu.fi}}
          \and
          V. Fallah Ramazani\inst{1,2}
          \and
          E. Lindfors\inst{1,3}
          \and
          P. Goldoni\inst{4}
          \and
          J. Becerra Gonz\'alez\inst{5,6}
          \and
          J. A. Acosta Pulido\inst{5,6}
          \and
          R. Clavero \inst{5,6}
          \and        
          J. Otero-Santos \inst{5,6,7}
          \and
          T. Pursimo\inst{8}
          \and
          S. Pita\inst{4}
          \and
          P. M. Kouch\inst{1,3}
          \and
          C. Boisson\inst{9}
          \and
          M. Backes\inst{10,11}
          \and
          G. Cotter\inst{12}
          \and
          F. D'Ammando\inst{13}
          \and 
          E. Kasai\inst{10}
          }

    \institute{Finnish Centre for Astronomy with ESO (FINCA), Quantum, Vesilinnantie 5, FI-20014 University of Turku, Finland
             \and
                Aalto University Metsähovi Radio Observatory, Metsähovintie 114, 02540, Kylmälä, Finland
               \and
               Department of Physics and Astronomy, University of Turku, Finland
              \and
               Universit\'e Paris Cit\'e, CNRS, CEA, Astroparticule et Cosmologie, F-75013 Paris, France
               \and
               Universidad de La Laguna (ULL), Departamento de Astrofísica, E-38206 La Laguna, Tenerife, Spain
               \and
               Instituto de Astrofísica de Canarias (IAC), E-38200 La Laguna, Santa Cruz de Tenerife, Spain
               \and
               Instituto de Astrof\'isica de Andalucía (CSIC), Glorieta de la Astronomía s/n, 18008 Granada, Spain
               \and
               Nordic Optical Telescope, Apartado 474, E-38700 Santa Cruz de La Palma, Spain
               \and
               Laboratoire Univers et Th{\'e}ories, Observatoire de Paris, Université PSL, Université Paris Cité, CNRS, F-92190 Meudon, France
               \and
               Department of Physics, Chemistry \& Material Science, University of Namibia, Private Bag 13301, Windhoek, Namibia
               \and
               Centre for Space Research, North-West University, Potchefstroom 2520, South Africa
               \and
               Oxford Astrophysics, University of Oxford, Denys Wilkinson Building, Keble Road, Oxford, OX1 3RH, United Kingdom
               \and
               INAF - Istituto di Radioastronomia, Via Gobetti 101, I-40129 Bologna, Italy
             }

   \date{Received September 15, 2023; accepted September 16, 2023}

 
  \abstract
   {}
   {Direct redshift determination of BL Lac objects is highly challenging as the emission in the optical and near-infrared (NIR) bands is largely dominated by the non-thermal emission from the relativistic jet that points very close to our line of sight. Therefore, their optical spectra often show no emission lines from the host galaxy. In this work, we aim to overcome this difficulty by attempting to detect the host galaxy and derive redshift constraints based on assumptions on the galaxy magnitude ("imaging redshifts").} 
   {Imaging redshifts are derived by obtaining deep optical images under good seeing conditions, so that it is possible to detect the host galaxy as weak extension of the point-like source. We then derive the imaging redshift by using the host galaxy as a standard candle using two different methods.}
   {We determine imaging redshift for 9 out of 17 blazars that we observed as part of this program. The redshift range of these targets is 0.28-0.60 and the two methods used to derive the redshift give very consistent results within the uncertainties. We also performed a detailed comparison of the imaging redshifts with those obtained by other methods, like direct spectroscopic constraints or looking for groups of galaxies close to the blazar. We show that the constraints from different methods are consistent and that for example in the case of J2156.0+1818, which is the most distant source for which we detect the host galaxy, combining the three constraints narrows down the redshift to $0.63<z<0.71$. This makes the source interesting for future studies of extragalactic background light in the Cherenkov Telescope Array Observatory era.}
   {}

   \maketitle


\section{Introduction}

Blazars are active galactic nuclei (AGN) in which relativistic jets, launched by supermassive black holes, point very close to our line of sight. They are the most numerous extragalactic gamma-ray \citep[E = 100 keV -- 100 GeV,][]{2020ApJ...892..105A} and very-high-energy \citep[VHE, E = 100 GeV -- 100 TeV,][]{2008ICRC....3.1341W} gamma-ray sources and candidate sources of very high energy astrophysical neutrinos detected by the IceCube Neutrino Observatory \citep{2018Sci...361.1378I}.

Blazars are classified into flat spectrum radio quasars \mbox{(FSRQs)} and BL Lacertae (BL Lac)  objects. The distinctive feature between the two classes is the presence of broad (few thousand km/s) and luminous ($\ge$ 10$^{42}$ erg s$^{-1}$) emission lines in the optical spectra of the former and their weakness or absence in the latter. The limit is usually set at an equivalent width EW of 5 \AA~\citep{1995PASP..107..803U} but it is known that EW emission lines of BL Lacs can sometimes get larger than that \citep[see e.g.] []{Stick91}. 
One challenge in studies of the many key questions of blazar physics is the difficulty of obtaining redshifts from their nearly featureless, continuum-dominated spectra.
High \mbox{signal-to-noise} spectra are needed to detect those weak lines and, therefore, a large fraction of BL Lacs ($\sim 59\%$) have no firmly established redshifts still today \citep{2015Ap&SS.357...75M}.

For BL Lac objects, it is in many cases easier to detect the host galaxy via deep optical and NIR imaging than the weak emission and absorption lines. It was shown by \citet {Sbar05} that the distribution of BL Lac host galaxy absolute magnitudes $M_R$ is almost Gaussian with an average value of -22.9 and $\sigma=0.5$. Therefore, BL Lac host galaxies can be used as "standard candles" to estimate their distances. This result was later confirmed by \cite{Shaw13}, albeit with slightly different mean absolute magnitude. Due to the width of the host galaxy luminosity distribution this redshift estimate has a 1-$\sigma$ error of at least 0.05 at $z$=0.3 or about 16\%, while spectroscopic redshift errors are regularly smaller than 0.1\%.

Even if this method is less accurate than spectroscopic determination, it still  gives an estimate of the distance of the source. In addition, it has been shown that for the sources where the host galaxy is detected in imaging, the spectroscopic observations are more likely to detect lines for more accurate redshift determination \citep{Nil03, Gol21,2024arXiv240107911D}. Furthermore, the accurate determination of the host galaxy magnitude from imaging observations also allows us to estimate the mass of the supermassive black hole (M$_{\mathrm {BH}}$), a parameter that has been suggested to significantly correlate with the shape of the emitted spectral energy distribution (SED) and its bolometric luminosity \citep{Ghis11}.

In the next few years a new facility, the Cherenkov Telescope Array Observatory (CTAO), will become operational with a northern site in the Canary Islands (Spain) and a southern site in the Atacama desert (Chile). It will have a 20 GeV - 300 TeV energy range and a sensitivity approximately 10 times better than the current facilities in this energy range. Blazars are the main extragalactic targets for CTAO, which is expected to detect hundreds of blazars according to current estimates, thus opening the possibility of population studies with a significant sample size \citep{2019scta.book.....C}. The planning of observations by the CTAO Consortium is currently ongoing. It is therefore of great importance to start acquiring redshifts and properties of the host galaxies for a large fraction of the AGN sources detected with {\it Fermi}-LAT that are likely to be detected with CTAO. There is, therefore, a redshift measuring campaign from which three papers have recently been published. \citet{Gol21} contains spectroscopic measurements of 19 BL Lac objects, resulting in determination of 11 new high-confidence redshifts, one tentative redshift and three lower limits to redshift. In \citet{Kas23}, 25 high-quality spectra are presented, resulting in 14 confirmed redshifts. In \citet{2024arXiv240107911D}, 24 high quality spectra resulted in 12 confirmed redshifts, one tentative redshift and two lower limits.

In this paper, we present I-band imaging of a sample of 17 blazars. The sample selection is summarised in Section 2 and the deep optical imaging observations in Section 3. The analyses to derive the imaging redshifts are explained in Section 4 and the results of these analyses in Section 5. The discussion of how the derived constraints on redshift of the sources compare with redshifts derived by other methods is presented in Section 6 and summary and conclusions are presented in Section 7. Throughout this paper we use
the cosmology $H_0$ = 70 km s$^{-1}$ Mpc$^{-1}$, $\Omega_{\Lambda}$ = 0.7 and $\Omega_{M}$ = 0.3.

\section{Sample selection}   

This paper is part of a larger redshift determination program and therefore the initial step of the sample selection was performed as in \citet{Gol21}, \citet{Kas23} and \citet{2024arXiv240107911D}. {\it Fermi}-LAT has detected more than 3130 active galactic nuclei \citep[4FGL catalogue;][]{2020ApJS..247...33A}. The sources that are detected above 10\,GeV by {\it Fermi}-LAT \citep[3FHL][]{2017ApJS..232...18A} have good potential to be detected also with CTAO. 3FHL contains 1040 BL~Lacs and Blazar candidates of an unknown type (BCUs) and initially for 64\% of the sources the redshifts are unknown. The 3FHL spectra, that typically represent the average flux state of the source, are extrapolated to the VHE $\gamma$-ray band, but adding cut-off at 3\,TeV. Absorption by the extragalactic background light (EBL) has also to be accounted for, but to account for it correctly, we would obviously need the redshift of the sources. At the time of the selection, we did not have spectroscopic redshifts for the sources and therefore used $z$=0.3 for all sources. The EBL model from \citet{Dom11} was used to calculate the gamma-ray absorption. The Monte Carlo simulations of the detectability with CTAO were performed  with the Gammapy software{\footnote{\url{https://gammapy.org/}} and the publicly-accessible CTAO performance files. The sources that are detectable at 5$\sigma$ in 30\,hours of observations with CTAO and have unknown redshift were selected for this redshift program. The full sample consists of 165 sources.

In this paper we present deep I-band imaging observations of a sub-sample of 17 sources for which we could not find evidence of extension or deep imaging observations from the literature. This program started in 2016, so for some sources spectroscopic redshifts have been obtained, some of them as follow-up of these imaging observations. In that case, we re-evaluated the detectability with CTAO using the actual redshift. The observation times for CTAO to detect these sources are given in Table~1.

\begin{table*}
\caption{\label{tab_sample}General properties of the sample.}        
\centering  
\setlength{\tabcolsep}{0.55em}
\begin{tabular}{lccccccc}          
\hline 
\multicolumn{1}{c}{(1)} & (2) & (3) & (4) & (5) & (6) & (7) \\
\multicolumn{1}{c}{\multirow{2}{*}{Source name}}  & \multirow{2}{*}{3FHL Name}&RA   &Dec 	& \multirow{2}{*}{z} & \multirow{2}{*}{ref} & CTA Exp.\\ 
& & J2000 & J2000 & &  & (hours)\\
\hline  
\object{GB6 J0045+2127} & J0045.3+2127 & 00 45 19.30 & 21 27 40.10 & 0.4253         & 1 &  2.5  \\
\object{GB6 J0148+5202} & J0148.2+5201 & 01 48 20.25 & 52 02 05.80 & 0.437	        & 1 &  42.0 \\
\object{GB6 J0905+1358} & J0905.5+1357 & 09 05 34.99 & 13 58 06.30 & 0.2239,0.644$^*$	& 1 &  8.1,20.5\\
\object{Ton 396       } & J0915.9+2933 & 09 15 52.40 & 29 33 24.04 &                &   &  15.5 \\
\object{RBS 1040	}   & J1150.5+4154 & 11 50 34.76 & 41 54 40.09 &     	        &   &  14.8 \\
\object{1RXS J154604.6+081912} & J1546.1+0818 &	15 46 04.30	& 08 19 14.00 &	>0.513	& 2  &  63.6 \\
\object{PG 1553+113	}   & J1555.7+1111 & 15 55 43.04 & 11 11 24.36        & 0.43	& 3  &  0.1  \\
\object{1RXS J181118.3+034109} & J1811.3+0341 &	18 11 18.09	& 03 41 13.90 &	        &    & 11.8 \\
\object{NVSS J184425+154646	}  & J1844.4+1546 &	18 44 25.34	& 15 46 46.00 &	 	    &    & 47.7 \\
\object{1RXS J193109.5+093714} & J1931.1+0937 &	19 31 09.23	& 09 37 16.40 &	>0.476	& 4   & 1.8  \\
\object{1RXS J193320.3+072616} & J1933.3+0726 &	19 33 20.30	& 07 26 16.00 &	        &    & 5.7  \\
\object{1RXS J194246.3+103339} & J1942.7+1033 &	19 42 47.48	& 10 33 27.80 &	     	&    & 0.9  \\
\object{RX J2030.8+1935} & J2031.0+1936	      & 20 30 57.13 & 19 36 12.91 & 0.3665	& 5   & 35.0 \\
\object{RX J2156.0+1818} & J2156.0+1818	      &	21 56 01.63	& 18 18 39.20 &	>0.6347	& 6   & 31.3 \\
\object{NVSS J224604+154437} & J2245.9+1545 & 22 46 04.99 & 15 44 37.50	  &	0.5966	& 5  &  65.1 \\
\object{1RXS J224753.3+441321} & J2247.9+4413 &	22 47 53.21	& 44 13 15.30 &	    	&    & 23.9 \\
\object{1RXS J230437.1+370506} & J2304.7+3705 &	23 04 36.80	& 37 05 07.00 &  	    &    & 27.5 \\
\hline
\end{tabular}
\tablefoot{Columns: (1) source name, (2) source name in 3FHL catalogue, (3) right ascension, (4) declination,  (5) assumed redshift,  (6) redshift reference, (7) CTA exposure time for the detection of the source at VHE gamma rays using the redshift or lower limit in the table or z = 0.3 if no redshift is available.

$^*$ See section \ref{ind_target} for details about the redshift values.

Redshift references: (1) \cite{Pai20}, (2) \cite{Ahu19} (3) \cite{John19} (4) \cite{Shaw13} (5) \cite{Kas23} (6) \citet{2024arXiv240107911D}.
}
\end{table*}

\section{Observations}

The observations were carried out with the Nordic Optical Telescope (NOT){\footnote{The Nordic Optical Telescope is located at Observatorio del Roque de los Muchachos, La Palma, Canary Islands, Spain. Its primary mirror diameter is 2.56\,m.}}. The objects of our sample were observed by NOT between 2016 and 2019 using the Alhambra Faint Object Spectrograph and Camera\footnote{\url{http://www.not.iac.es/instruments/alfosc/}} (ALFOSC) in \textit{I}-band\footnote{\url{http://www.not.iac.es/instruments/filters/curves/png/12.png}}. This instrument has a field of view of 6.4 $\times$ 6.4 arcmin with pixel scale of 0.21 arcsec/pix., gain 0.15 e$^-$/ADU and readout noise 4.3 e$^-$. The dark current is negligible {due to liquid nitrogen cooling (0.03 e$^-$/pix/h at
$\rm T_{operation}=-120^\circ$\,C)}. The observations were performed in I-band during good seeing conditions to maximise the probability of detecting the host galaxy. The details of the observations are given in Table \ref{tab_obs}.

\begin{table*}
\caption{\label{tab_obs}Logbook of the observations.}          
\centering  
\begin{tabular}{lccccccc}          
\hline 
\multicolumn{1}{c}{(1)} & (2) & (3) & (4) & (5) & (6) & (7)\\
\multicolumn{1}{c}{\multirow{2}{*}{3FHL name}}  & \multirow{2}{*}{Date} &  \multirow{2}{*}{N} & Exp. & FWHM & Airmass & Ap\\ 
& &  & (s) &  (arcsec)  &  range& mag\\
\hline  
J0045.3+2127	&	2019-08-30	&	            20	& 2900	& 0.59 & 1.01-1.07 & 16.57\\
J0148.2+5201	&	2019-08-30	&	            10	& 1900	& 0.69 & 1.09-1.11 & 16.92\\
J0905.5+1357	&	2019-03-02, 2019-03-01	&	27	& 4680  & 0.70 & 1.03-1.07 & 16.35\\
J0915.9+2933	&	2018-01-10	&	            81	& 2430	& 0.74 & 1.05-1.20 & 15.40\\
J1150.5+4154	&	2019-03-01, 2019-05-06	&	43	& 5130	& 0.76 & 1.04-1.06 & 16.04\\
J1546.1+0818	&	2019-05-07, 2019-06-17	&	36	& 3600  & 0.69 & 1.06-1.08 & 16.64\\
J1555.7+1111	&	2019-05-06, 2019-06-17	&	63	& 315   & 0.67 & 1.06-1.15 & 12.94\\
J1811.3+0341	&	2019-08-29	&	             7	& 2900	& 0.90 & 1.10-1.13 & 16.25\\
J1844.4+1546	&	2019-08-29	&	            10	& 2250  & 0.86 & 1.08-1.16 & 16.46\\
J1931.1+0937	&	2016-10-01	&	            10	& 2000  & 0.78 & 1.20-1.34 & 16.14\\
J1933.3+0726	&	2017-06-19	&           	12	& 2400  & 0.57 & 1.07-1.09 & 16.58\\
J1942.7+1033	&	2016-10-09	&	            12	& 1400  & 0.51 & 1.32-1.51 & 15.61\\
J2031.0+1936	&	2019-05-07, 2019-06-20	&	16	& 3200	& 0.72 & 1.04-1.21 & 17.24\\
J2156.0+1818	&	2019-08-30	&	            20	& 2850  & 0.70 & 1.02-1.05 & 16.79\\
J2245.9+1545	&	2019-08-30	&	             6	& 3000  & 0.74 & 1.02-1.03 & 18.15\\
J2247.9+4413	&	2017-07-26	&           	12	& 2400  & 0.59 & 1.29-1.48 & 16.93\\
J2304.7+3705	&	2017-08-11	&	            12	& 2400  & 1.00 & 1.01-1.03 & 17.17\\
\hline
\end{tabular}
\tablefoot{Columns:  (1) source name in 3FHL catalogue, (2) date(s) of observation, (3) number of images combined, (4) total exposure time in seconds, (5) seeing (FWHM) in arcsec {which is calculated by fitting a Gaussian profile to a star close to the target}, (6) airmass range during the exposures and (7) I-band magnitude through a 5 arcsec diameter aperture. The error is 0.02 mag.}
\end{table*}

\section{Analysis}

The main goal of the analysis is to obtain the host galaxy luminosities and effective radii, which can then be used to derive the redshift using two methods described in more detail below. For this purpose we fitted two-dimensional models to the observed light distribution. 

Prior to the fitting, the images were reduced in the standard way of bias subtraction and flat-fielding, after which the individual images were registered using suitable stars in the field and summed. Calibration of the summed images was achieved through the stars with known magnitudes in the blazar fields. We first matched the positions of isolated stars in the field with positions in the PAN-STARRS survey catalogue \citep{2016arXiv161205560C}. The number of matched stars varied from 7 to 77 per field. We then obtained their $i$ and $z$ -band aperture magnitudes from PAN-STARRS and transformed them to Cousins I-band via the formula\footnote{\url{https://live-sdss4org-dr12.pantheonsite.io/algorithms/sdssUBVRITransform/\#Lupton200}}

$$
I = i - 0.3780 (i-z) - 0.3974.
$$
After performing aperture photometry to the stars in our images, we derived the zero point for each field as the average difference between our and the transformed PAN-STARRS magnitudes. The rms scatter of the magnitude differences was generally found to be very small, about 0.01-0.02 mag per field, even for fields with the most numerous matches, thus showing that our calibration is accurate.

After calibration the local background near the blazar value was determined from 5-7 empty sky regions within $\sim$ 10 arcsec of the target and subtracted. A final check was made to ensure that there are no subtle slopes in the background. In the few cases a slope was found, it was corrected by fitting a plane to the pure sky emission around the target and subtracting the fit. Then we determined if there were any stars, galaxies, blemishes, etc. overlapping the blazar image and masked any affected pixels out of the fit. Finally, we visually determined the radius beyond which no blazar light {(from host galaxy and nucleus)} was visually detectable and restricted the fit inside this area.

Our model for the light distribution of blazars consists of two components, the unresolved AGN (hereafter the "core") and the host galaxy. Three parameters, ($x_c$,$y_c$) position plus magnitude $m_c$ are sufficient to fully describe the core, whereas 7 parameters are needed for the host galaxy: position ($x_g,y_g$, assumed to be the same as the core), magnitude $m_g$, effective radius $r_{\rm eff}$, ellipticity $\epsilon$, major axis position angle $\theta$ and the S\`ersic index $n$, which determines how centrally concentrated the light distribution is {by assuming that the light follows the S\`ersic profile}

\begin{equation}
    I(R)=I_e \exp\left\{ -b_n\left[ \left( \frac{R}{R_e}\right) ^{1/n} -1\right] \right\}.
\end{equation}
{where $R_e$ is the half-light radius, $I_e$ is the intensity at $R_e$, and $b_n$ is a coefficient which depends on the S\`ersic index $n$.}

The models were convolved with a point-spread function (PSF) obtained from a high S/N star as close as possible to the target. We prefer an empirical PSF over an analytical one since, according to our experience, the former better represents the complexities of the PSF, especially for images like ours that are made by summing data taken over a course of several hours. Empirical PSFs contain observational noise, but this noise can be made insignificant by using a sufficiently bright star for the PSF. As discussed in e.g. \cite{Nil03}, the PSF shape depends on the position on the ALFOSC CCD. We did not attempt to model this, but rather included it in our noise model (see below).

For the fitting we used the downhill simplex method to find the set of parameters that minimises the chi square between the data and model.  To compute the chi square we assumed the noise in each pixel to consist of three components, summed in quadrature to calculate the total noise per pixel: photon noise, readout noise and PSF error. The first two can be computed from the effective gain and the effective read out noise. The PSF error was assumed to be equal to a Gaussian function centred on the target with a width equal to the FWHM of the image and an amplitude of $A$ times the peak value of the target. This noise component is significant only close to the centre of the target due to the quick tapering of the Gaussian function. A reasonable range for $A$ is 0.01-0.1. We checked through simulations how the fit results depend on the choice of $A$. We found that, within the above range of $A$, the fit results are not biased in any significant way by the choice of $A$. However, the scatter of retrieved host galaxy parameters increases with increasing $A$. This happens because as $A$ increases, the PSF noise becomes a more significant, and eventually dominant, noise source at the centre of the target. The central pixels contribute less and less to the total chi square sum, making the chi square space "flatter," thus increasing the scatter in the retrieved parameters. We found $A = 0.03$ to provide a reasonable estimate of the PSF errors for our data and adopted this value for all targets.

The fits were done in two stages: we first fitted a pure AGN model ($x_c$,$y_c$,$m_c$) to the image to fix the position and to look for any excess light in the residual image indicative of a host galaxy. In practice, if the core-subtracted image showed no residuals beyond PSF errors, we marked the target as unresolved, since experience has shown that no host galaxy is detected by the fit in this case (typically $r_{\rm eff}$ tends towards zero or infinity). In case any residuals were detected, we moved to stage 2 by fixing the core and the host to the position obtained in stage 1. Since the host galaxies are typically very small and the unresolved core complicates the host characterisation, we did not attempt a detailed analysis of the host morphology. Therefore, we fixed ellipticity $\epsilon$ and position angle $\theta$ to 0 and let only the host magnitude $m_g$ and effective radius $r_{\rm eff}$ to vary freely. Thus the second stage had 3 free parameters as well ($m_c$,$m_g$,$r_{\rm eff}$). Stage 2 was done for two S\`ersic indices, $n=1$ and $n=4$ with the first representing a disk-dominated galaxy and the latter a bulge-dominated galaxy. We find that $n=4$ consistently gives a better fit to these observations, and, except where noted, the results presented are for the $n=4$ case.

We used two different methods to estimate the blazar redshift. Firstly, the results by \cite{Sbar05} and \cite{Shaw13} show that the BL Lac host galaxy luminosities are confined to a relatively narrow range of  $\pm$ 0.5 mag. The former obtained average luminosity $M_R$ = -22.9, whereas the latter  $M_R$ = -22.5. The latter also discuss two likely reasons for the difference in these results. Firstly, the \cite{Shaw13} sample may contain fainter hosts simply for the reason that the more luminous hosts were already discovered by the time of their survey since they are easier to detect. The other reason may be a  difference in the mixture of low energy peaked (LBL) and high energy peaked (HBL) BL Lacs between the samples. For our analysis we assume that $M_R = -22.7$ and that $R -I = 0.62$ at z=0. The latter value was determined by integrating an elliptical galaxy template spectrum of \cite{Mannu01} over the R-band and NOT I-band band-passes. Since the transformation from apparent to absolute magnitudes depends on  redshift $z$, we have to find $z$ iteratively. Because we also want to obtain an estimate of the uncertainty of $z$, we repeat the redshift iteration 1000 times each with different inputs. Thus we had two loops, the inner loop performing the $z$ iteration and the outer loop repeating this iteration 1000 times.

For each $z$ iteration we first drew the observed I-band magnitude of the host galaxy $m_I$  from a Gaussian distribution with a mean and $\sigma$ equal to the observed I-band  magnitude and its error, respectively, and corrected it for Galactic absorption. Similarly, we drew a value for the host galaxy luminosity $M_R$ from a Gaussian distribution with mean = -22.7 and $\sigma$ = 0.5 and transformed that to $M_I$. Then we made a first guess of the redshift $z=z_0$ and computed the K-correction $K(z_0)$ and evolution correction $e(z_0)$ corresponding to this guess. The K-correction was computed from a fit to the correction in \cite{Fuku95}. For the evolution correction we used $E(z) = 0.84 \times z$. This relation was derived by using the  Pegase 3 code \citep{2019A&A...623A.143F} with a single starburst 10 Gyr ago and passive evolution thereafter. After this we derived the distance modulus 

\begin{equation}
m - M = m_I - M_I - K(z_0) + E(z_0)
\end{equation}
{ and the corresponding redshift $z_1$ from }
\begin{equation}
m -M = 5 \log{D_L} + 5.0  ,
\end{equation}
{ where $D_L$ is the luminosity distance in parsecs. Since there is no closed
from solution between $z_1$ and $D_L$, we interpolated the $z_1$ corresponding
to $D_L$ from a pre-computed table.}

Then $z_1$ was used to calculate $K(z_1)$ and $E(z_1)$ to derive $z_2$. This calculation was iteratively repeated until it stabilised, which typically happened after $\sim 5$ iterations and the resulting $z$ was stored. After the outer loop was finished we computed the mean and standard deviation of the 1000 $z$ estimates.  

In the second method we assume that our host galaxies follow the Kormendy relation, i.e.the projection of the fundamental plane to the $R_{\rm eff} - \langle \mu \rangle$ axes of elliptical galaxies, where $R_{\rm eff}$ is the effective radius in kpc and $\langle \mu \rangle$ is the average surface brightness inside the effective radius. This method uses both the host galaxy brightness and the effective radius as input unlike the first method that uses only the brightness.

We used the i-band Kormendy relation derived in \cite{2020Ap&SS.365..142S} for the brightest cluster galaxies (BCGs)

\begin{equation}
\label{samir}
\langle \mu \rangle = (3.75 \pm 0.04) * \log (R_{\rm eff} /1\, {\rm kpc}) + (16.40 \pm 0.04).
\end{equation}
Although blazar host galaxies are typically not found in dense clusters \citep[e.g][]{1997ApJ...480..547W}, their luminosities are very similar to the BCGs. The redshift is found as follows: we start with a relatively low redshift $z=0.05$ and compute the corresponding effective radius $R_{\rm eff}$ in kpc from the observed effective radius $r_{\rm eff}$. Next, we compute  $\langle \mu \rangle$ from the observed host galaxy magnitude $m$ using and $r_{\rm eff}$

\begin{equation}
\langle \mu \rangle = m + 2.5 \log(2 \pi (r_{\rm eff} / 1 {\rm \, arcsec})^2) - K(z) + E(z) - 10\log(1+z).
\end{equation}
This will result in a point in the $R_{\rm eff} - \langle \mu \rangle$ plane that lies below and left of the line defined by Eq. \ref{samir}. Increasing the test redshift will cause the point to move up and towards the right, until at certain $z = z_{\rm best}$ it eventually  cuts through the line defined by Eq. \ref{samir} (see Fig. \ref{kormendyfig}). The redshift at which this happens is then adopted as the redshift of the blazar. We repeated this process 1000 times drawing each time the host magnitude, host effective radius, Galactic absorption and the slope and intercept of Eq. \ref{samir} from their respective distributions and adopted the standard deviation of the $z_{\rm best}$ as the error for the $z$.

\begin{figure}
\centering
\includegraphics[width=0.49\textwidth]{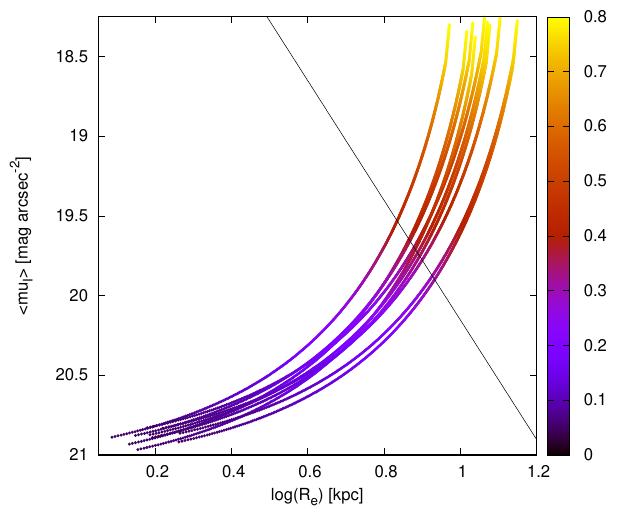}
\caption{\label{kormendyfig}Example of 10 runs with the "Kormendy" method for J0045.3+2127. Each run starts at z=0.05 in the lower
left corner, moves up and right as $z$ is increased and ends
at z=0.8 as indicated by color. The line indicates the Kormendy relation of Eq. \ref{samir}.}
\end{figure}

\begin{figure*}
\centering
\includegraphics[width=0.8\textwidth]{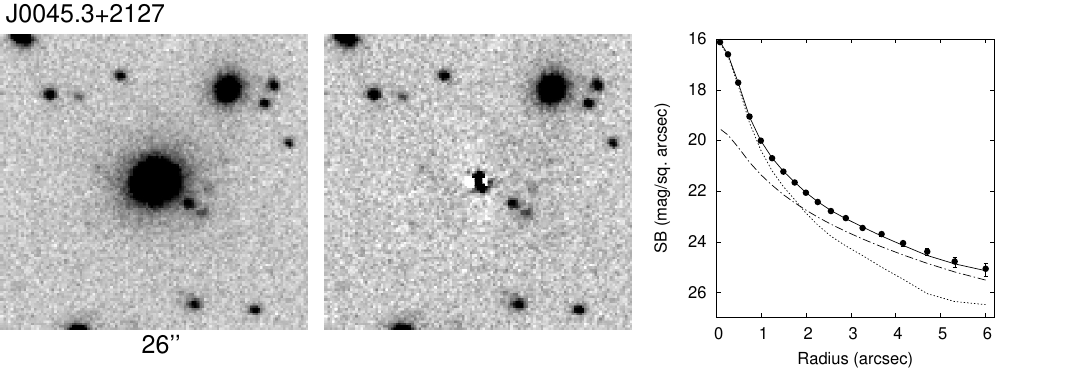}
\includegraphics[width=0.8\textwidth]{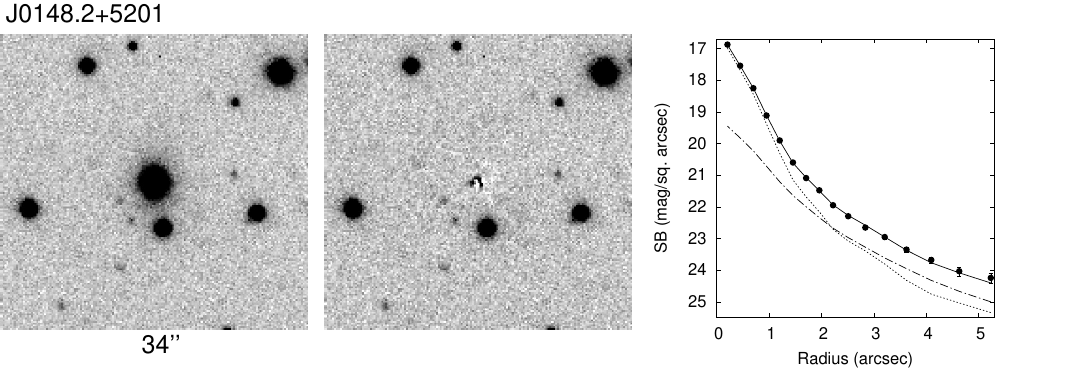}
\includegraphics[width=0.8\textwidth]{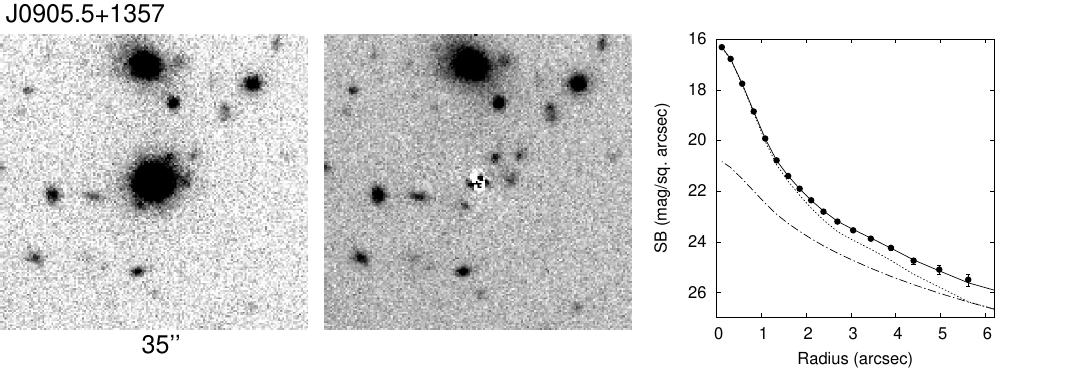}
\includegraphics[width=0.8\textwidth]{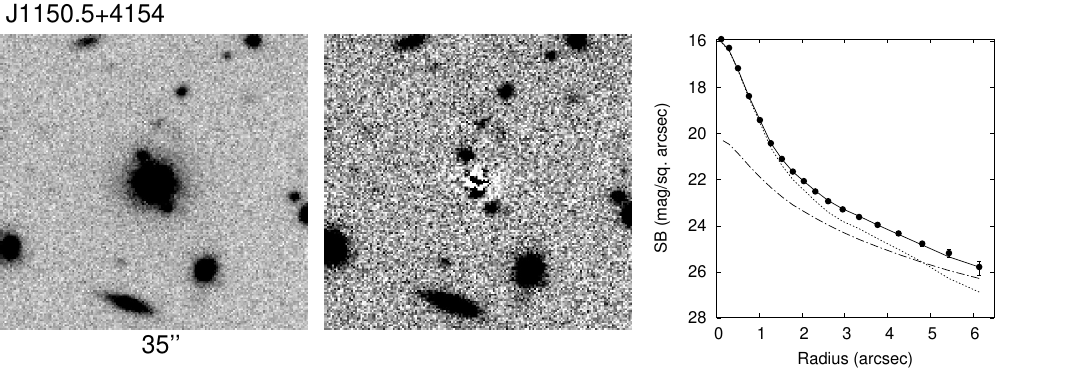}
\caption{\label{fig1}Results of the AGN + host galaxy decomposition. {\it Left panel} : Greyscale image of the target. North is up and East is to the left. The field of view is given under the figure. {\it Middle panel}: the same image after subtracting the model. {\it Right panel}: Radial surface brightness profiles of the target (filled symbols), model (solid line), AGN nucleus (dashed line) and host galaxy (dot dashed line).}    
\end{figure*}

\setcounter{figure}{1}

\begin{figure*}
\centering
\includegraphics[width=0.8\textwidth]{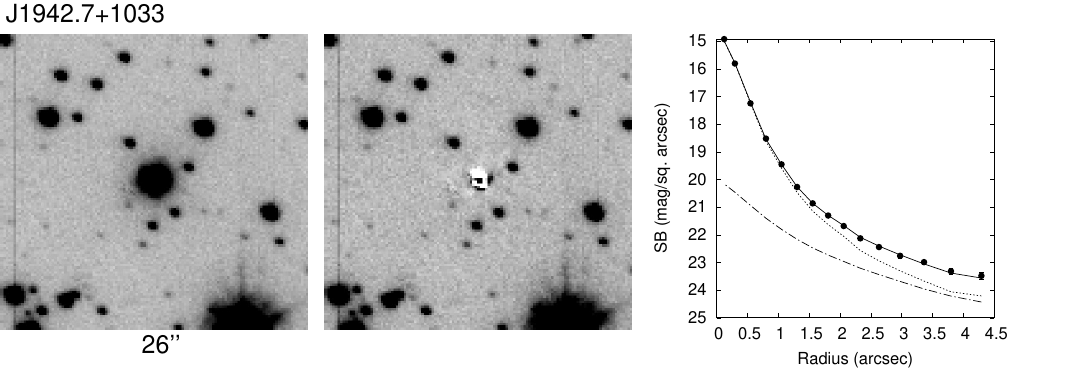}
\includegraphics[width=0.8\textwidth]{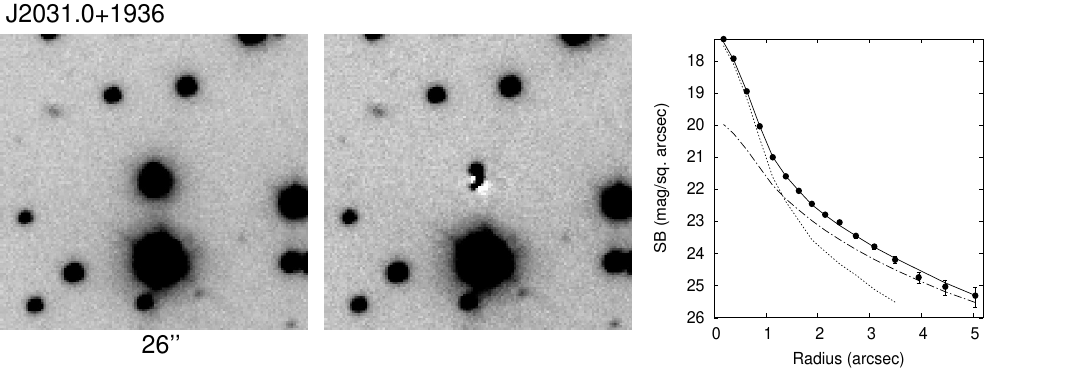}
\includegraphics[width=0.8\textwidth]{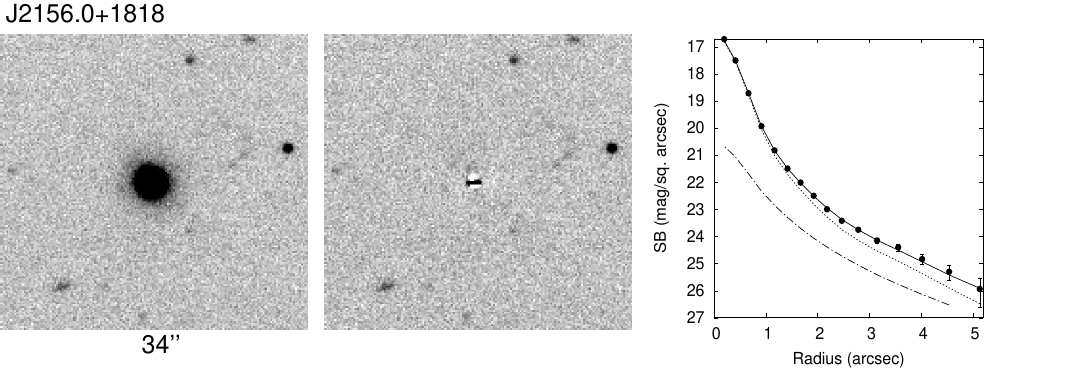}
\includegraphics[width=0.8\textwidth]{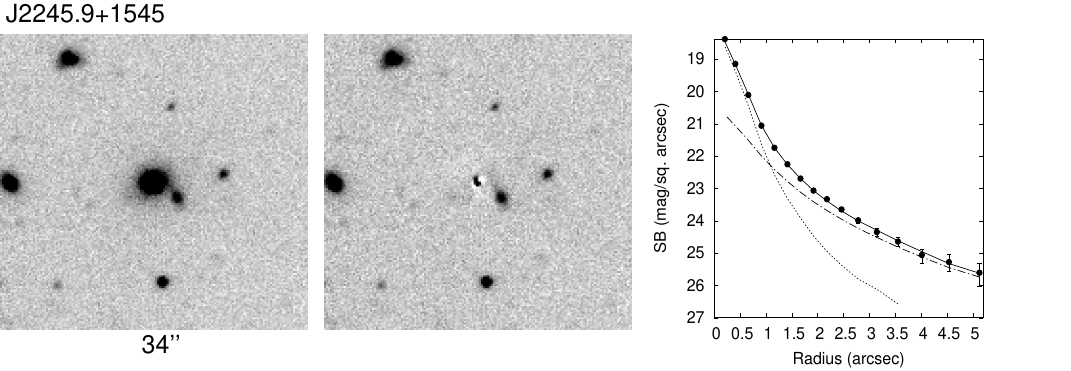}
\caption{--Continued}
\end{figure*}

\setcounter{figure}{1}

\begin{figure*}
\centering
\includegraphics[width=0.8\textwidth]{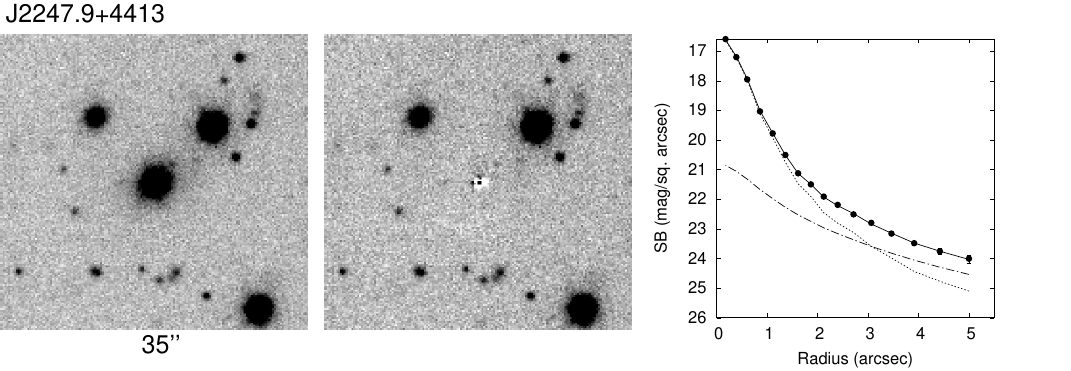}
\caption{--Continued}
\end{figure*}

\begin{table*}
\caption{\label{tab_results}Results of the { host galaxy analysis}.  }   
\centering  
\begin{tabular}{lcccccccc}          
\hline 
\multicolumn{1}{c}{(1)} & (2) & (3) & (4) & (5) & (6) & (7)\\  
\multicolumn{1}{c}{\multirow{2}{*}{3FHL name}}& $I_{nucleus}$  & $I_{galaxy}$  & $r_{\rm eff}$  & $A_{\text{I}}$ & \multirow{2}{*}{$z_1$} & \multirow{2}{*}{$z_2$}\\
& (mag) & (mag) & (arcsec) & (mag) & & &\\
\hline  

J0045.3+2127 & 16.72 $\pm$ 0.02 & 17.99 $\pm$ 0.05 & 1.7 $\pm$ 0.2 & 0.018 & 0.34 $\pm$ 0.07 & 0.34 $\pm$ 0.04 \\

J0148.2+5201 & 17.14 $\pm$ 0.03 & 18.31 $\pm$ 0.04 & 0.8 $\pm$ 0.2 & 0.025 & 0.38 $\pm$ 0.07 & 0.34 $\pm$ 0.07\\

J0905.5+1357 & 16.36 $\pm$ 0.02 & 19.03 $\pm$ 0.07 & 1.6 $\pm$ 0.5 & 0.053 & 0.50 $\pm$ 0.09 & 0.53 $\pm$ 0.12 \\

{ J0915.9+2933} & { 15.37 $\pm$ 0.02} & { 17.66 $\pm$ 0.08} & { 1.0 $\pm$ 0.2} & { 0.037} & { $\dagger$} &\\ 

J1150.5+4154 & 16.05 $\pm$ 0.02 & 18.78 $\pm$ 0.11 & 2.9 $\pm$ 0.7 & 0.027 & 0.46 $\pm$ 0.09 & 0.54 $\pm$ 0.10 \\

{ J1546.1+0818} & { 16.54 $\pm$ 0.03} & { $>$ 20.0} & & { 0.076} & { $>$ 0.5} &\\ 

{ J1555.7+1111} & { 13.16 $\pm$ 0.03} & { $>$ 17.7} & & { 0.078} & { $>$ 0.2} &\\

{ J1811.3+0341} & { 16.15 $\pm$ 0.02} & { $>$ 18.9} & & { 0.289} & { $>$ 0.3} &\\

{ J1844.4+1546} & { 16.35 $\pm$ 0.01} & { $>$ 19.3} & & { 0.687} & { $>$ 0.3} &\\

{ J1931.1+0937} & { 16.14 $\pm$ 0.02} & { $>$ 18.6} & & { 1.024} & { $>$ 0.2} &\\

{ J1933.3+0726} & { 16.72 $\pm$ 0.03} & { $\dagger$} &  & { 0.452} & { $\dagger$} &\\

J1942.7+1033 & 15.61 $\pm$ 0.02 & 18.01 $\pm$ 0.12 & 3.1 $\pm$ 0.6 & 0.578 & 0.28 $\pm$ 0.06 & 0.31 $\pm$ 0.06 \\

J2031.0+1936 & 17.47 $\pm$ 0.02 & 18.45 $\pm$ 0.07 & 1.8 $\pm$ 0.3 & 0.144 & 0.39 $\pm$ 0.08 & 0.40 $\pm$ 0.06 \\

J2156.0+1818 & 16.79 $\pm$ 0.02 & 19.63 $\pm$ 0.12 & 1.4 $\pm$ 0.7 & 0.131 & 0.60 $\pm$ 0.11 & 0.62 $\pm$ 0.14 \\

J2245.9+1545 & 18.45 $\pm$ 0.02 & 18.74 $\pm$ 0.11 & 2.4 $\pm$ 0.5 & 0.116 & 0.44 $\pm$ 0.09 & 0.49 $\pm$ 0.08 \\

J2247.9+4413 & 16.99 $\pm$ 0.02 & 18.32 $\pm$ 0.16 & 3.8 $\pm$ 0.9 & 0.370 & 0.34 $\pm$ 0.07 & 0.41 $\pm$ 0.09 \\

{ J2304.7+3705} & { 17.16 $\pm$ 0.03} & { $>$ 19.2} & & { 0.222} & { $>$ 0.35} &\\

\hline
\end{tabular}
\tablefoot{Columns: (1) source name in 3FHL catalogue, (2) magnitude of the AGN nucleus, (3) magnitude of the host galaxy, (4) effective radius of the host galaxy, (5) galactic absorption from \citet{2011ApJ...737..103S}, (6) redshift derived from the host galaxy brightness and (7) redshift derived through the Kormendy relation.\\ { $\dagger$ Not estimated due to morphological anomalies}.}
\end{table*}

\section{Results}

The host galaxies of 9 out of 17 objects have been detected and the results of the  AGN + host galaxy decomposition for these sources are shown in Figure~\ref{fig1}. The measured magnitudes and effective radii of the galaxies along with the resulting imaging redshifts using the two methods described in previous section are presented in Table \ref{tab_results}. In all cases the imaging redshifts derived with the two methods agree very well within the uncertainties.

{ We also estimated lower limits of $z$ for the unresolved targets by adding
an elliptical galaxy with $M_R = -21.7$, i.e. 1 mag fainter than the average value,
and r$_{\rm eff}$ = 8 kpc to the observed image at increasing redshifts.
We then determined the redshift at which the host galaxy became visually undetectable.
Since it is not easy to objectively determine what "not visible" means, we applied fairly
conservative judgement and the resulting lower limits can also be considered very
conservative.}

The observations presented here had a good success rate of ($\sim$ 53\%), similar to the previous work by \citep{Nil03}, where 100 BL Lacertae objects were imaged through the R-band. It is well known that it is easier to detect the host galaxy when the core is in a low state. We have indeed tried to optimize the timing of our imaging and spectroscopic \citep[see][]{Gol21} observations to lower states of the blazars, when we have this information available. To understand if the successful detections were preferentially the result of observations in a low state and to see if there is room for improvement for the ones we did not detect, we investigated the Zwicky Transient Facility (ZTF) light curves of the 12 of our sources that were observed with NOT in the ZTF era (i.e. after beginning of 2018). In Appendix A, we show these light curves. As can be seen, most of the 12 sources were observed in rather moderate activity state, i.e. we managed to avoid the brightest flares, but did not catch the minima of the light curves either. The three exceptions to this were 3FHL~J1555.7+1111 (PG~1553+113), which we observed in rather high state (and did not detect the host), 3FHL~J0905.5+1357 and 3FHL~J2031.0+1936, which we did observe in low state (and detected the host galaxy). Judging from the light curves, all four sources (3FHL~J0915.9+2933, 3FHL~J1546.1+0818, 3FHL~J1555.7+1111, 3FHL~J1844.4+1546), for which we have the ZTF data and did not detect the host should be re-observed during lower brightness state (see also Section 6).

\begin{figure*}
\centering
\includegraphics[width=0.8\textwidth]{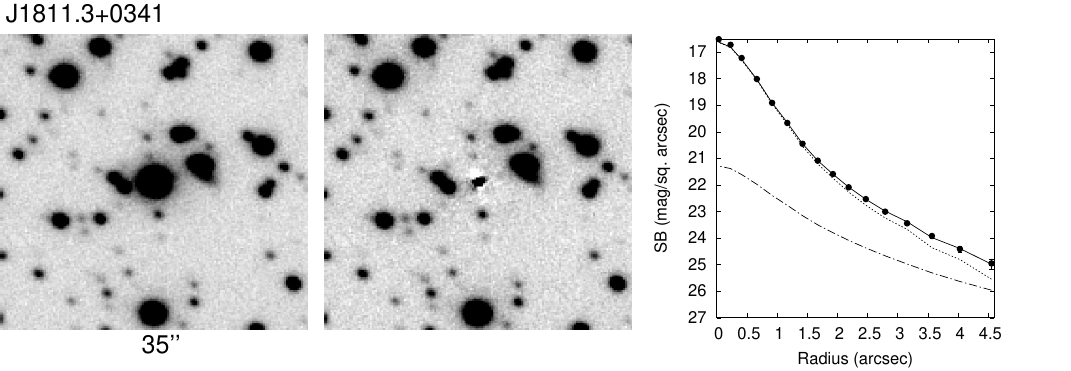}
\caption{Results of the AGN + host galaxy decomposition of J1811+0341. The panels are the same as in Fig. \ref{fig1}.
\label{1811proffig}}
\end{figure*}

Three unresolved targets deserve further comment. Firstly, we did not mark 3FHL~J1811.3+0341 (1RXS~J181118.3+034109) as resolved, although our analysis gives I$_{\rm host}$ = 19.22 $\pm$ 0.20 and r$_{\rm eff}$ = 1.7 $\pm$ 0.9 arcsec. As can be seen from Fig. \ref{1811proffig}, the surface brightness profile is just above the PSF and the host is very weak compared to the core. Furthermore, there are numerous objects overlapping with the blazar, making it very hard to find pixels unaffected by them. For these two reasons it is very difficult to reliably characterise the host galaxy and derive its redshift.

\begin{figure}
\includegraphics[width=0.23\textwidth]{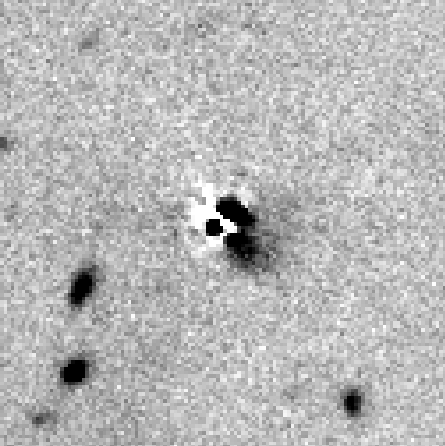}
\includegraphics[width=0.23\textwidth]{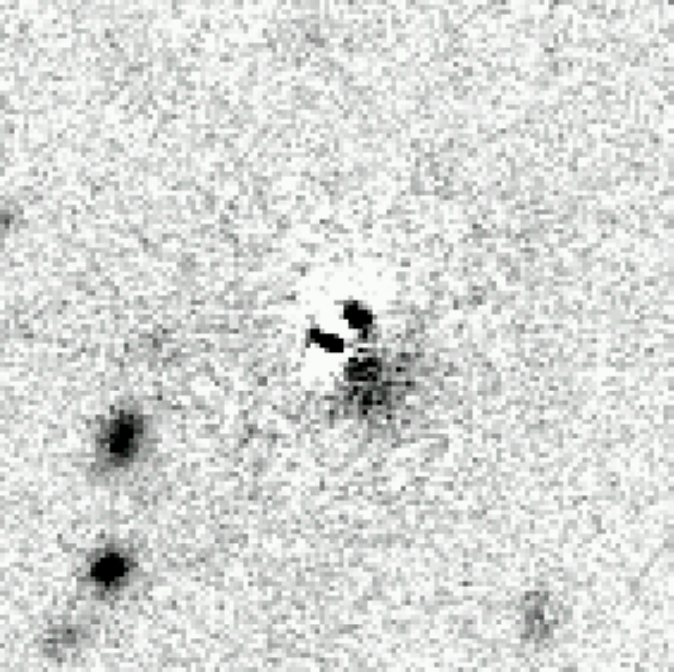}\\
\includegraphics[width=0.23\textwidth]{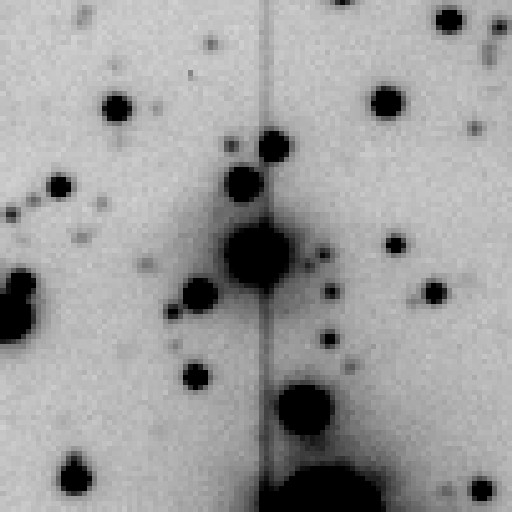}
\includegraphics[width=0.23\textwidth]{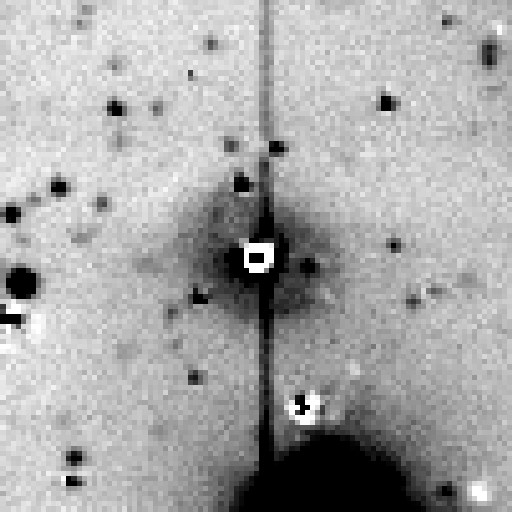}
\caption{{\em Upper row, left}: 3FHL~J0915.9+2933 after subtracting the { PSF
+ host galaxy model}. The field of view is 27$\times$27 arcsec. {\em Upper row, right}: the same target from \cite{Nil03}. {\em Lower row, left} : a 21 $\times$ 21 arcsec field around 3FHLJ~1933.3+0726. {\em Lower row, right} : The same field after subtracting the PSF from bright sources in the field and from the blazar. \label{0915sub}}
\end{figure}

Secondly, some earlier studies of 3FHL~J0915.9+2933 have concluded that this object is not extended \citep{Abr91,Nil03} while \citet{Mei10} found it extended with i$_{gal}$ $\sim$ 18.45. We see weak excess around the centre after subtracting the PSF from the position of 3FHL~J0915.9+2933 (Fig \ref{0915sub}).  Fitting a AGN core + host galaxy model yields  { i = 17.66 $\pm$ 0.08 and $r_{\rm eff}$ = 1.0 $\pm$ 0.2 arcsec}. There is a faint extension SW of centre (Fig. \ref{0915sub}, upper row), also visible in the residual image of \cite{Nil03} { and some excess
emission E of the target. Due to these anomalies, we do not attempt to estimate
the redshift of this source. The residuals after subtracting the
core correspond to i = 18.20, consistent with \citet{Mei10}}.

Thirdly, 3FHL~J1933.3+0726 shows very atypical features for a blazar (Fig. \ref{0915sub}, lower row). This 
target is clearly resolved, but the host galaxy type
cannot be assigned since, unlike in all other targets, where
the host galaxy fades smoothly into the background 
noise, the nebulosity around 3FHL~J1933.3+0726 exhibits
a relatively sharp edge, giving an impression of  
spiral galaxy seen face-on. The numerous targets overlapping
the blazar and a CCD bleeding feature running over
the core of J1933.3+0726 further complicate the analysis.
We used an iterative scheme in order to mitigate
the effect of overlapping stars. We first performed PSF photometry
on all bright targets near 3FHL~J1933.3+0726 and subtracted them.
Then we ran the model fit and subtracted the blazar model from
the original image. The PSF photometry was repeated again on this
subtracted image after which model fitting was done again after
subtracting the blazar model. After three iterations the fit
results stabilized. This iteration was done using $n=1$ and $n=4$
and notably this is the only target where  the $n=1$ model gave
a better fit to the data. However, significant residuals are still
present in both cases. Given the peculiar morphology, the high number of overlapping targets and instrumental blemishes we did not attempt to derive a redshift for this target.

\begin{figure}
\centering
    \includegraphics[width=0.45\textwidth]{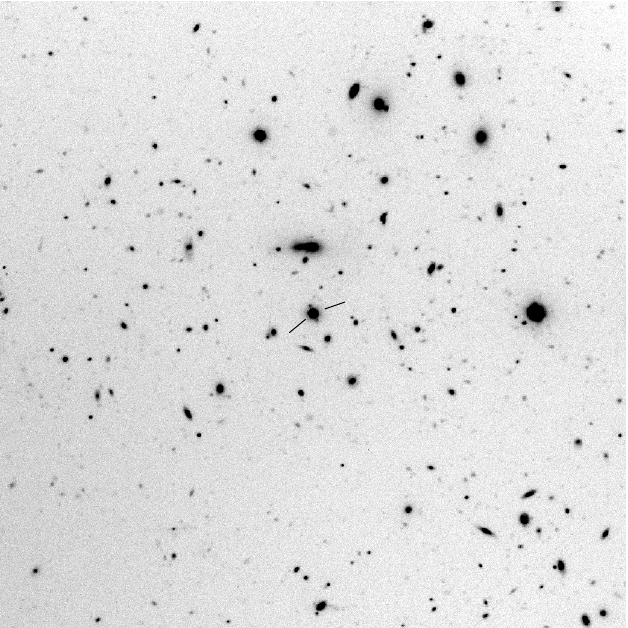}
    \caption{\label{1150field}4.3 $\times$ 4.3 arcmin field around 3FHL~J1150.5+4154. The blazar is indicated by tick marks.}
\end{figure}

In one of the fields, 3FHL~J1150.5+4154, the galaxies around the blazar seem to be distorted in a manner characteristic to gravitational lensing (Fig. \ref{1150field}). We made a simple study to see if we could establish this quantitatively by looking for an excess of galaxies with their sky projected major axes perpendicular to the line connecting the galaxy to 3FHL~J1150.5+4154. We first extracted all targets brighter than certain magnitude $m_c$ and within a certain radius $r$ of 3FHL~J1150.5+4154 using the Source Extractor \citep{1996A&AS..117..393B} and recorded their positions, ellipticities $\epsilon$ and position angles. Then we
calculated $\phi$, the angle between the target major axis and the line
connecting the target to 3FHL~J1150.5+4154, and plotted
$\phi$ as a function of $\epsilon$. In case of strong distortions
by lensing it is expected that at low $\epsilon$, where the targets are mostly stars, $\phi$ is
independent on $\epsilon$, but as $\epsilon$ gets higher the
$\phi$ values are more concentrated around $90$ degrees. We tested
different values of $m_c$ and $r$, but in all cases the distribution of $\phi$ remained uniform, even at high ellipticities. We thus found no evidence of preferred orientation
of galaxies around 3FHL~J1150.5+4154.

\section{Discussion}

In this section, we compare the redshift constraints obtained in this work with other redshift constraints. As the redshifts we determined in this work with the two methods agree in all cases very well within the uncertainties, in the following comparisons we only report z$_1$ for simplicity.

\begin{table}[tb]
\caption{Comparison of host galaxy based redshift estimates to spectroscopically measured redshifts. The upper part lists earlier imaging results, whereas the lower part refers to the 
results presented in this paper.}
    \label{zcompa}
    \centering
    \begin{tabular}{ccccc}
    \hline
    (1) & (2) & (3) & (4) & (5)\\
    Target  &  z$_{\rm phot}$ & ref & z$_{\rm spec}$ & ref\\
    \hline
    RGBJ0115+253 & 0.35 $\pm$ 0.05 & 2 & 0.376 & 3\\ 
    RGBJ0202+088 & 0.55 $\pm$ 0.05 & 2 & 0.629 & 10\\
    RGBJ0227+020 & 0.45 $\pm$ 0.05 & 2 & 0.458 & 4\\  
    RGBJ0250+172 & 0.25 $\pm$ 0.05 & 2 & 0.243 & 3\\
    RGBJ0505+042 & 0.35 $\pm$ 0.05 & 2 & 0.424 & 11\\
    PKS0735+178  & 0.45 $\pm$ 0.06 & 6 & 0.424 & 7\\  
    RGBJ0757+099 & 0.30 $\pm$ 0.05 & 2 & 0.266 & 3\\  
    RGBJ1415+485 & 0.50 $\pm$ 0.05 & 2 & 0.496 & 5\\
    \hline
    3FHLJ0045.3+2127 & 0.34 $\pm$ 0.07 & 1 & 0.4253 & 8\\
    3FHLJ0148.2+5201 & 0.38 $\pm$ 0.07 & 1 & 0.437  & 8\\
    3FHLJ0905.5+1357 & 0.50 $\pm$ 0.09 & 1 & 0.644$^*$  & 8\\
    3FHLJ2031.0+1936 & 0.39 $\pm$ 0.08 & 1 & 0.3665 & 9\\
    3FHLJ2245.9+1545 & 0.44 $\pm$ 0.09 & 1 & 0.5966 & 9\\
    \hline
    \end{tabular}
    \tablefoot{$^*$ See section \ref{ind_target} for details about the redshift values.
    
    References: 
    (1) This work,
    (2) \cite{Nil03}, 
    (3) \cite{2022MNRAS.515.4810L},
    (4) \cite{2005AJ....129..559S},
    (5) SDSS DR3,
    (6) \cite{2012A&A...547A...1N},
    (7) \cite{2018MNRAS.473.5154M},
    (8) \cite{Pai20},{ $^1$}
    (9) \citet{Kas23},
    (10) \cite{2012ApJ...748...49S}
    (11) \cite{2014A&A...565A..12P}
    }
\end{table}

In 5 cases out of 9 where we detected the host galaxy, we have also the spectroscopic redshift from \citet{Pai20} or \citet{Kas23}. Table \ref{zcompa} lists those cases together with earlier estimates from the literature. The photometric and spectroscopic redshifts are compared in Fig. \ref{zpred}. This figure shows that the host galaxy method is able to provide a reasonable estimate of the blazar redshift at least up to $z$= 0.5. Beyond that, the redshifts seem to be underestimated by the host galaxy method, but there are only three data points, so it is unclear  how significant this offset is. Such a deviation is expected since at redshifts beyond 0.5 we are likely to detect only the most luminous host galaxies, brighter than the assumed $M_R$ = -22.7.

\begin{figure}
\centering
\includegraphics[width=0.45\textwidth]{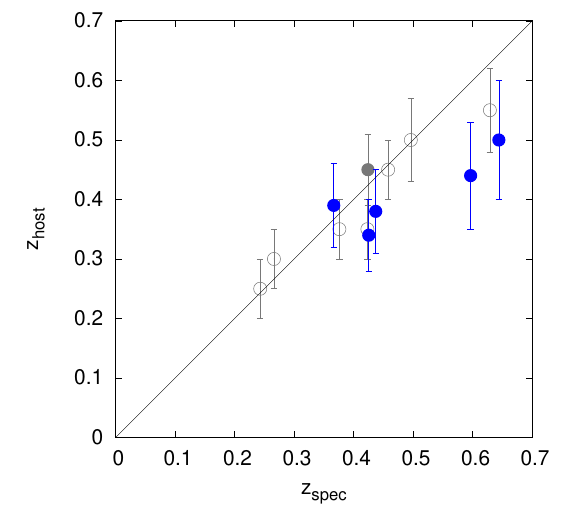}
\caption{\label{zpred}Comparison of redshifts obtained spectroscopically and the host galaxy method. Grey symbols show earlier results and blue symbols the results obtained in this paper. Open symbols denote R-band detections, filled symbols I-band (see Tab. \ref{zcompa} and the text for details). The line shows perfect correspondence.
}
\end{figure}

This can in fact be seen from Table \ref{bhmass} where we have computed the R-band host galaxy luminosity and estimated the central black hole masses of the 5 targets with a spectroscopic redshift. To estimate the black hole masses we first transformed our K- and evolution corrected I-band absolute magnitudes to the R-band using $M_R = M_I + 0.62$. We then used two different bulge-luminosity - BH mass relationships, the $M_R$ - $M_{BH}$ correlations of \cite{Gra07} and \cite{Jia11} to compute the black hole masses. The two estimates agreed to within 0.1 in dex. In Table \ref{bhmass} we give the average of the two results. As can be seen from the table, in 4 out of 5 cases the luminosity of the host galaxy is brighter than the average luminosity $M_R$ = -22.7, in three cases even more than the $\sigma=0.5$ of the host galaxy absolute magnitude distribution.

On the other hand, some targets in our 17 target sample have been imaged before with negative results. Thus, one might argue that because of this, the bright host galaxies have already been detected and we should be finding on average fainter host galaxies. \cite{Shaw13} discussed this as one possible explanation why they derived fainter average magnitude ($M_{R}=-22.5$) for their host galaxies compared to earlier work by \cite{Sbar05} ($M_{R}=-22.9$). In our case, none of the five targets had previous imaging observations and the sample is too small to make any statement to support one or the other average magnitude. But it is certainly interesting that in this very small sample, the very luminous host galaxies are "over-represented". We encourage further deep imaging observations of high-synchrotron-peaked sources with known spectroscopic redshift to further investigate this issue.

\begin{table*}
    \caption{\label{bhmass}R-band host galaxy luminosity and estimated the central black hole masses.}
    \centering  
    \begin{tabular}{cccccc}
    \hline
    (1) & (2) & (3) & (4) & (5) & (6)\\
         3FHL name & $z$ & $M_R$ & r$_{\rm eff}$ (kpc)& $\log(M_{BH}/M_{\odot})$ & ref \\
         \hline
         J0045.3+2127 &  0.4253    & $-23.3 \pm 0.1$  & 9.5   $\pm$ 1.1 & 8.7    & 1\\
         J0148.2+5201 &  0.437     & $-23.1 \pm 0.1$  & 4.5   $\pm$ 1.1 & 8.6    & 1\\
         \multirow{2}{*}{J0905.5+1357$\biggl\{$} &  0.2239  & -20.7 $\pm$ 0.1 & 5.7 $\pm$ 1.8& 8.0 & \multirow{2}{*}{$\biggl\}$1}\\
                     &  0.644     & -23.4 $\pm$ 0.2  & 11.0   $\pm$ 3.5 & 8.7    &\\
         J2031.0+1936 &  0.3665    & -22.6 $\pm$ 0.1  & 9.1   $\pm$ 1.5 & 8.5    & 2\\
         J2245.9+1545 & 0.5966     & -23.6 $\pm$ 0.3
         & 16.0 $\pm$ 3.3  & 8.8  & 2\\
         \hline
    \end{tabular}
    \tablefoot{Columns: (1) source name in 3FHL catalogue, (2) redshift, (3) absolute magnitudes, (4) effective radii, (5) black hole mass, (6) redshift reference. 
    
    References: (1) \cite{Pai20}, (2) \citet{Kas23}. }
\end{table*}
In addition to direct spectroscopic constraints on redshifts and the host galaxy imaging discussed in this paper, it is also possible to obtain constraints on the blazar redshift from the spectroscopic redshifts of galaxies close to the blazar. This method has been used for several blazar previously: PKS 0447-439 \citep[$z$=0.343,][]{Mur15}, PKS 1424+240 \citep[$z$=0.601,][]{Rov16}, PG~1553+113 \citep[$z$=0.433,][]{John19} which is also in our sample (3FHL~J1555.7+1111), and RGB~2243+203 \citep[$z$=0.528,][]{Ros19}.  In the case of PKS 1424+240, direct spectroscopic observations have later confirmed the redshift \citep{Pai17,2024arXiv240107911D}. We did not perform dedicated spectroscopy of the companions, but eight of the blazars in our sample are within the SDSS footprint and we searched the SDSS database to obtain the redshifts of the galaxies within 7.5 arcmin of the target blazar. We selected this radius, because at z=0.3 this radius corresponds to approximately 2\,Mpc, which is a typical search radius for companion galaxies \citep[see e.g.][]{Massaro20}. According to mock sample simulations of \cite{Massaro19a,Massaro19b}, if there are two companions at the same redshift (within $z<0.05$) within 1.0\,Mpc from the blazar and$/$or 4 companions within 2\,Mpc, the probability that the blazar is associated to that group is higher than 95\%. Their studies extend only to redshift of 0.15 and are therefore not directly applicable to our objects that are at higher redshifts. Nevertheless, we search for companion galaxies within these two radii. The redshifts of the companion galaxies, with their distance to the central blazar are reported in Appendix~B. Figure \ref{histograms} shows the two best examples of companion galaxy redshift distribution. Interestingly, we find for seven objects a group of four or more galaxies within 7.5 arcmin from the blazar. The only object for which we do not find any such group is 3FHL~J1555.7+1111 which may be linked to the weakness of the galaxies composing the group found by \citet{John19}. In the case of 3FHL~J1546.1+0818 the group is at redshift $<0.1$ and therefore likely to be a foreground group. More detailed comparison for individual objects can be found below.

We also investigated our images searching for faint galaxies that could be observed with dedicated spectroscopic observations. However, many of the objects that are not within SDSS footprint are in very crowded galactic fields (3FHL~J1811.3+0341, 3FHL~J1931.1+0937, 3FHL~J1933.3+0726, 3FHL~J1942.7+1033 and 3FHL~J2031.0+1936) and identifying faint galaxies in these fields was challenging even from our deep images.

\begin{figure}
\centering
\includegraphics[width=0.45\textwidth]{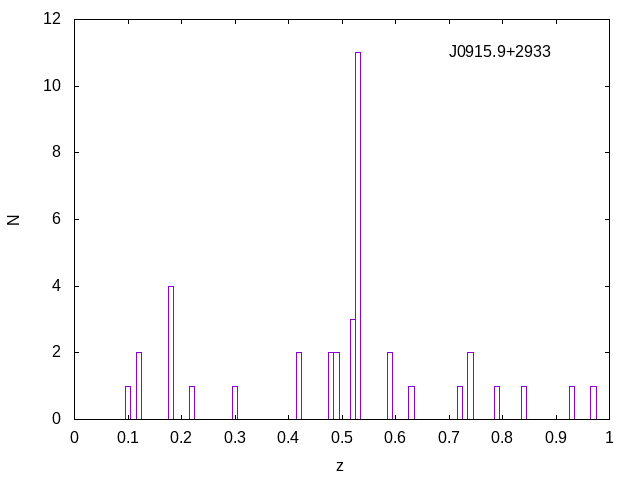}
\includegraphics[width=0.45\textwidth]{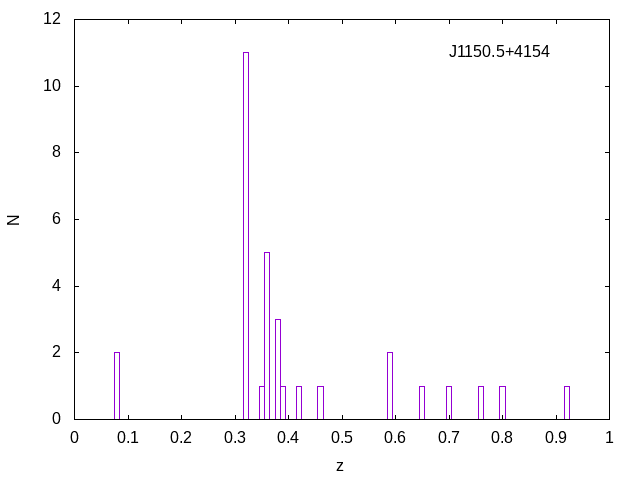}
\caption{
\label{histograms}
Histograms of companion galaxy redshifts. The redshift bins are 0.01 and the histograms are shown only for the sources that have five or more galaxies in one of the redshift bin.}
\end{figure}

\subsection{Notes on individual sources}\label{ind_target}

\begin{itemize}
\item 3FHL~J0045.3+2127: Spectroscopic redshift~$z$=0.4253 \citep{Pai20}. As our imaging redshift is $0.34\pm0.07$, the host galaxy of this object is more luminous than average host galaxies of BL Lac objects. In SDSS we also find 5 galaxies within 0.6\,Mpc (assuming z=0.42) from the source, suggesting that the source is a member of rather rich group of galaxies.

\item 3FHL~J0148.2+5201: Spectroscopic redshift~$z$=0.437 \citep{Pai20}. As our imaging redshift is $0.38\pm0.07$, the host galaxy is on the brightest edge of the host galaxy brightness distribution. The source is outside of the SDSS footprint.

\item 3FHL~J0905.5+1357: Two spectroscopic observations separated by about one year interval led to the detection of one emission line. If this line is interpreted as [OIII]b, then~$z$=0.2239, while if interpreted as [OII] then~$z$=0.644 \citep{Pai20}. Our imaging result is $0.50\pm0.09$. The calculated absolute magnitudes of the host galaxy assuming the two spectroscopic values show that if at $z=$0.2239, the galaxy is very faint ($M_R$=-20.9), and we therefore consider $z=$0.644 the more likely redshift, even if this indicates a very luminous host galaxy $M_R=-23.4$. We note that even if outside the range suggested by \cite{Sbar05} ($M_R=-22.9$, $\sigma=0.5$), the derived $M_R$ at z=0.644 is as bright as the host galaxy of J2245.9+1545. In SDSS we find no galaxies at redshift close to 0.22 within 5 arcsec from the blazar, but there are three galaxies with redshift of $z=0.644\pm0.05$, one of which is the closest companion of the target. We note that there is a group of faint galaxies well visible in our image and for which SDSS does not give a spectroscopic redshift as they are too faint. In summary, z=0.644 is the likely redshift of this blazar and to establish it we suggest deep low state spectroscopic observations. Also spectroscopic observations of the faint galaxies in the field could be used to establish the presence of a group and hence constrain the redshift and characterise the environment of the BL Lac.

\item 3FHL~J0915.9+2933: Several featureless high S/N spectra exist \citep{Shaw13,Ahu19,Pai17}. Also the determination of the imaging redshift was not successful (see Section 5). However, the field has been extensively investigated by \citet{Witt06} in their selection of clusters on the basis of weak lensing shear. They found a cluster candidate, DLCS J0916.0+2931, centreed 2.5' arcmin from  the position of the blazar, extending about ten arc minutes in the North-South direction. Subsequent Chandra observations detected two extended X-ray emission regions 3.2 arcmin North and 5.9 arcmin South of the blazar and a possible extended region 27 arcsec from the blazar. The Northern and Southern X-ray sub-clusters are both at z=0.53 from Keck optical \citep{Witt06} and XMM observations \citep{Desh17}. The central sub-cluster could not be detected in XMM observations due to confusion with the strong blazar emission \citep{Desh17} and optical spectroscopy of its galaxies was not performed. The presence of this extended cluster in the field argues for 3FHL~J0915.9+2933 to be a high redshift gamma-ray blazar. In SDSS the closest galaxy has redshift 0.489 (assuming this redshift, it is only 200 kpc from the blazar), but within 2 arc minutes (about 800 kpc at those distances) there are also two galaxies at 0.53 like the DLCS cluster, and another one at 0.487. When we consider the whole area of 7.5 arcmin, there are in total 11 galaxies at z$\sim$0.53. Therefore, from companion galaxy data alone, we cannot conclude the redshift. We simply note that both "candidate" redshifts from the candidate companion galaxy data are rather high. As mentioned in Section~5, the imaging observations were not done in particularly low state of the AGN and deep imaging in lower state could allow us to detect the host galaxy and estimate the redshift. However, in the case of this source, the large uncertainties of the imaging redshift would most likely contain all three possible redshift values and therefore a high S/N spectrum obtained during photometric low state would be needed to finally determine the redshift.

\item 3FHL~J1150.5+4154: \citet{White00} claimed z=1.01 for this source, but in literature can be found also several featureless spectra \citep{Shaw13,Ahu19,Pai20}. The last one has high signal to noise and was obtained in a relatively low state. In this work, we derive an imaging redshift of $0.46\pm0.09$, incompatible with the high redshift suggested by \cite{White00}. Our result is consistent with the lower limit set by \cite{Pai20}, z$>0.25$. In SDSS, 11 nearby galaxies within 3.5 arc minutes (about 1 Mpc at z=0.32) have spectroscopic redshift $z$ $\sim$ 0.322-0.328. Our imaging redshift suggests that the blazar is located at higher redshift than this group or that the host galaxy of the blazar is under-luminous $M_R=-21.8$, which we also cannot exclude. However, as discussed in Section 5, there is visual indication of gravitational lensing in this field, so even if we were not able to confirm that, we note that the imaging redshift would be also consistent with the group being the gravitational lens.

\item 3FHL~J1546.1+0818: A spectroscopic lower limit $z \ge$ 0.513 \citep{Ahu19} has been obtained by SDSS. We observed it using Lick/Shane (\citet{Kas23}) but we obtained only a low S/N spectrum. The implied high redshift is in agreement with our non-detection of the host galaxy. As mentioned above already, the only group of 4 galaxies in this field has redshift of $z\sim0.07$ and is clearly a foreground group, especially given that the spectroscopic lower limit indicates that 2\,Mpc at the redshift of the blazar is smaller than 5 arcmin, while two of the galaxies of this group have a distance to the blazar greater than 6 arcmin. Within this radius there is a group of three galaxies with redshifts 0.55-0.57. The closest companion has redshift of 0.513, exactly at the redshift of the lower limit from spectroscopy and at a distance of $\sim$285 kpc. In summary, we cannot conclude the redshift, but {our lower limit $z>0.5$ supports the redshift estimations above.}

\item 3FHL~J1555.7+1111: (a.k.a. PG~1553+113) is a very well known VHE $\gamma$-ray emitting blazar. There have been multiple attempts of imaging \citep{Scarpa00, Treves07} and high S/N spectra \citep{Sba06,hessVLT1553,Lan14} without success. Therefore, it is not surprising that our imaging did not detect any extension of the object, especially as it was taken in a rather high state of the source (see light curves in Appendix A). For this blazar a spectroscopic lower limit z$\ge$ 0.4131 was reported \citep{Dan10}. Furthermore it is reported to be a member of a galaxy group at z=0.433 \citep{John19,Dor22} and this gives the strongest constraint on the redshift. \citet{Dor22} reported the spectroscopic redshift z=0.408 - 0.436 (95\% confidence interval), but these galaxies are too faint for SDSS spectroscopy and therefore do not show up in our search for companions. Indeed in our deep image, we do detect many faint galaxies.

\item 3FHL~J1811.3+0341: We have previously obtained two featureless medium S/N spectra  \citep[Lick/Kast,][]{Kas23}. As discussed in Section 5, we see some extended emission, but it is too weak to determine the redshift {and resulted in $z>0.3$}. From the light curve it is evident that the timing of the imaging observations was more favourable than the timing of the spectra and therefore deeper spectroscopy should be obtained during a lower optical state to determine the spectroscopic redshift.

\item {3FHL~J1844.4+1546: We derived lower limit of $z>0.3$. We are not aware of previous constraints on the redshift of this source.} 

\item 3FHL~J1931.1+0937: The spectroscopic lower limit $z \ge$ 0.476 \citep{Shaw13} is in agreement with {our lower limit $z>0.2$.} 

\item 3FHL~J1933.3+0726: As discussed in Section~5, the extension we see is unlikely to be the host galaxy of the source. \citet{Kas23} obtained a high S/N Keck/ESI spectrum of this target, which turned out to be featureless, i.e. showed no lines from the extension we see. 3FHL~J1933.3+0726 was $\sim$ 1 mag brighter during the epoch of the spectrum than during host imaging and the low surface brightness ($\sim$ 22.4 mag / sq. arcsec) of the nebulosity are likely to be the two main factors contributing to the non-detection in the spectrum. As the object is not within the SDSS footprint, our study does not result in new information about the redshift of the target and it remains unknown.

\item 3FHL~J1942.7+1033: One featureless high S/N spectrum has been obtained with FORS/VLT \citep{Tsa05,Mase13}, but we clearly detect the host galaxy and obtain an imaging redshift of $z=0.28\pm0.06$. As the imaging observation was done before the ZTF era, we do not know if the central AGN was in a particularly low state during the imaging, but given the rather low redshift, the spectral lines should be also detectable if spectroscopy was performed in a low state.

\item 3FHL~J2031.0+1936: The spectroscopic redshift $z$=0.3665 \citep{Kas23,Pai23} is in good agreement with our imaging result $0.39\pm0.08$. This is in slight contradiction with the spectroscopic absolute magnitude of \citet{Kas23}, who reported the host galaxy to be quite luminous $M_R = -23.5 \pm 0.3$, whereas our results indicate $M_R = -22.6 \pm 0.1$. We reviewed the results by the \citet{Kas23}  and noticed that they obtain an observed magnitude r=17.7 for the source while ZTF obtains r=18.1 around this date. Correcting for this systematic difference, we would obtain $M_R = -23.1 \pm 0.3$, compatible with the imaging result. Our observations were performed during a low state of the source as was the Keck/ESI spectroscopic observation that resulted in the detection of the spectral lines and determination of the spectroscopic redshift.

\item 3FHL~J2156.0+1818: A spectroscopic lower limit $z \ge$ 0.6347 was obtained by \citet{2024arXiv240107911D}. Our imaging result is in agreement, and this is the most distant host galaxy in our sample for which we have significant detection of host galaxy. We derive z$=0.60\pm0.11$. There is a nearby galaxy in SDSS with z=0.6334 at corresponding distance of 1.16 Mpc, so very close to the spectroscopic lower limit, but at that distance it is likely not the origin of the absorber. Within the radius of 7.5 arcmin there is also one group of 4 galaxies with redshifts 0.51-0.53, but given the likely redshift of $>0.6$, all of these galaxies would be outside the 2\,Mpc radii and therefore not likely to be associated to the blazar.

\item 3FHL~J2245.9+1545: A spectroscopic redshift $z$=0.5965 was obtained by \citet{Kas23}, while the imaging redshift is $z=0.44\pm0.09$. So the host galaxy is very luminous, the most luminous in our sample. There is a group of four galaxies at $z=0.43$ and the galaxy closest to the blazar belongs to that group. There are no galaxies at the redshift similar to spectroscopic redshift, but given the high redshift, they could be too faint for SDSS. 

\item 3FHL~J2247.9+4413: Two featureless low S/N spectra can be found in the literature \citep{Shaw13,Mas15}. A featureless medium S/N  Lick/Kast spectrum was reported in \citet{Kas23}. Our imaging observations took place before the ZTF era, so we cannot evaluate if we did catch the sources in a particularly low state, but the imaging redshift of $z=0.34\pm0.07$ would indicate that with higher S/N, some lines should be detectable.

\item 3FHL~J2304.7+3705: A featureless low S/N spectrum was shown in \citep{Shaw13} and two featureless low S/N Lick/Kast spectra in \citet{Kas23}. We did not detect the host galaxy and the source is not within the SDSS footprint. {However, we estimated the lower limit $z>0.35$ for this target.}
\end{itemize}

\section{Summary and Conclusions}

In this work we present the results of a deep I-band imaging program of 17 blazars, whose $\gamma$-ray flux and spectral shape make them potential targets for the CTAO. At the time of our observations, the redshifts of the sources were unknown and the primary goal was to determine imaging redshifts for them. This was done using two different methods, one using the magnitude of the host galaxy and another using both magnitude and effective radii. In addition, we obtained redshifts of galaxies near our targets from the SDSS to further constrain the blazar redshifts.

We were able to detect the host galaxy in 9 out of 17 blazars. The imaging redshifts obtained with the two methods agree within errorbars. The derived imaging redshifts range from $0.28\pm0.06$ to $0.60\pm0.11$, with the latter (3FHL J2156.0+1818) being the most distant blazar for which the redshift has been constrained using this method. There were three targets where we detect the host galaxy, but direct spectroscopic observations have not detected any lines: 3FHL~J1150.5+4154, 3FHL~J1942.7+1033 and 3FHL~J2247.9+4413. The derived host galaxy magnitudes and parameters indicate that all of them have rather high nucleus to host galaxy ratio which could partially explain the non-detection of lines. Another explaining factor could be that the imaging observations were taken during lower state of the nucleus than the spectroscopic observations, but unfortunately in two of the cases the imaging observations were taken before the ZTF era and we could not evaluate this. The success of detecting the host galaxy, despite the high nucleus to host galaxy ratio, shows that in some of such cases the imaging redshift can indeed be the only viable way to get information about the source redshift.

We also showed that in the absence of a spectroscopically determined redshift it is still possible to obtain useful constraints for $z$ by combining deep imaging, archival spectroscopic data of the galaxy environment of the blazar and  spectroscopic lower limits. For instance, in the case of  3FHL J2156.0+1818 our imaging redshift $0.60\pm 0.11$ was also confirmed by spectroscopic observations and we found from SDSS that there is a nearby galaxy located at similar redshift. All three constraints are in good agreement and can be combined to narrow down the redshift of the object to the range $0.63466<z<0.71$, with the lower limit being a "hard" limit and the upper limit being a 1 sigma limit from the host galaxy. For 3FHL J0905.5+1357, we detect a host galaxy with a magnitude that clearly favours the higher of the two possible redshifts from \citet{Pai20}.

Finally, we note that in the case where we also had a spectroscopic redshift of the target and could therefore calculate the absolute magnitudes of the host galaxies, the detected hosts were brighter than what are typically detected for BL Lacs. Our sample is small and we cannot draw any conclusions from this, but this certainly serves as motivation for further studies of BL Lac host galaxies also in the future.

\begin{acknowledgements}
Based on observations made with the Nordic Optical Telescope, owned in collaboration by the University of Turku and Aarhus University, and operated jointly by Aarhus University, the University of Turku and the University of Oslo, representing Denmark, Finland and Norway, the University of Iceland and Stockholm University at the Observatorio del Roque de los Muchachos, La Palma, Spain, of the Instituto de Astrofisica de Canarias. The data presented here were obtained with ALFOSC, which is provided by the Instituto de Astrofisica de Andalucia (IAA) under a joint agreement with the University of Copenhagen and NOT.  Part of this work was supported by the German \emph{Deut\-sche For\-schungs\-ge\-mein\-schaft, DFG\/} project number Ts~17/2--1 and Academy of Finland projects 317636, 320045, 346071, and 322535. 

Funding for the SDSS and SDSS-II has been provided by the Alfred P. Sloan Foundation, the Participating Institutions, the National Science Foundation, the U.S. Department of Energy, the National Aeronautics and Space Administration, the Japanese Monbukagakusho, the Max Planck Society, and the Higher Education Funding Council for England. The SDSS Web Site is http://www.sdss.org/. The SDSS is managed by the Astrophysical Research Consortium for the Participating Institutions. The Participating Institutions are the American Museum of Natural History, Astrophysical Institute Potsdam, University of Basel, University of Cambridge, Case Western Reserve University, University of Chicago, Drexel University, Fermilab, the Institute for Advanced Study, the Japan Participation Group, Johns Hopkins University, the Joint Institute for Nuclear Astrophysics, the Kavli Institute for Particle Astrophysics and Cosmology, the Korean Scientist Group, the Chinese Academy of Sciences (LAMOST), Los Alamos National Laboratory, the Max-Planck-Institute for Astronomy (MPIA), the Max-Planck-Institute for Astrophysics (MPA), New Mexico State University, Ohio State University, University of Pittsburgh, University of Portsmouth, Princeton University, the United States Naval Observatory, and the University of Washington.

This work was conducted in the context of the CTAO Consortium.
\end{acknowledgements}

\bibliographystyle{aa}
\bibliography{ref.bib}

\begin{thebibliography}{59}
\expandafter\ifx\csname natexlab\endcsname\relax\def\natexlab#1{#1}\fi

\bibitem[{{Abdollahi} {et~al.}(2020){Abdollahi}, {Acero}, {Ackermann},
  {Ajello}, {Atwood}, {Axelsson}, {Baldini}, {Ballet}, {Barbiellini},
  {Bastieri}, {Becerra Gonzalez}, {Bellazzini}, {Berretta}, {Bissaldi},
  {Blandford}, {Bloom}, {Bonino}, {Bottacini}, {Brandt}, {Bregeon}, {Bruel},
  {Buehler}, {Burnett}, {Buson}, {Cameron}, {Caputo}, {Caraveo}, {Casandjian},
  {Castro}, {Cavazzuti}, {Charles}, {Chaty}, {Chen}, {Cheung}, {Chiaro},
  {Ciprini}, {Cohen-Tanugi}, {Cominsky}, {Coronado-Bl{\'a}zquez}, {Costantin},
  {Cuoco}, {Cutini}, {D'Ammando}, {DeKlotz}, {de la Torre Luque}, {de Palma},
  {Desai}, {Digel}, {Di Lalla}, {Di Mauro}, {Di Venere}, {Dom{\'\i}nguez},
  {Dumora}, {Fana Dirirsa}, {Fegan}, {Ferrara}, {Franckowiak}, {Fukazawa},
  {Funk}, {Fusco}, {Gargano}, {Gasparrini}, {Giglietto}, {Giommi}, {Giordano},
  {Giroletti}, {Glanzman}, {Green}, {Grenier}, {Griffin}, {Grondin}, {Grove},
  {Guiriec}, {Harding}, {Hayashi}, {Hays}, {Hewitt}, {Horan},
  {J{\'o}hannesson}, {Johnson}, {Kamae}, {Kerr}, {Kocevski}, {Kovac'evic'},
  {Kuss}, {Landriu}, {Larsson}, {Latronico}, {Lemoine-Goumard}, {Li},
  {Liodakis}, {Longo}, {Loparco}, {Lott}, {Lovellette}, {Lubrano}, {Madejski},
  {Maldera}, {Malyshev}, {Manfreda}, {Marchesini}, {Marcotulli},
  {Mart{\'\i}-Devesa}, {Martin}, {Massaro}, {Mazziotta}, {McEnery}, {Mereu},
  {Meyer}, {Michelson}, {Mirabal}, {Mizuno}, {Monzani}, {Morselli},
  {Moskalenko}, {Negro}, {Nuss}, {Ojha}, {Omodei}, {Orienti}, {Orlando},
  {Ormes}, {Palatiello}, {Paliya}, {Paneque}, {Pei}, {Pe{\~n}a-Herazo},
  {Perkins}, {Persic}, {Pesce-Rollins}, {Petrosian}, {Petrov}, {Piron}, {Poon},
  {Porter}, {Principe}, {Rain{\`o}}, {Rando}, {Razzano}, {Razzaque}, {Reimer},
  {Reimer}, {Remy}, {Reposeur}, {Romani}, {Saz Parkinson}, {Schinzel},
  {Serini}, {Sgr{\`o}}, {Siskind}, {Smith}, {Spandre}, {Spinelli}, {Strong},
  {Suson}, {Tajima}, {Takahashi}, {Tak}, {Thayer}, {Thompson}, {Tibaldo},
  {Torres}, {Torresi}, {Valverde}, {Van Klaveren}, {van Zyl}, {Wood},
  {Yassine}, \& {Zaharijas}}]{2020ApJS..247...33A}
{Abdollahi}, S., {Acero}, F., {Ackermann}, M., {et~al.} 2020, \apjs, 247, 33

\bibitem[{{Abraham} {et~al.}(1991){Abraham}, {McHardy}, \& {Crawford}}]{Abr91}
{Abraham}, R.~G., {McHardy}, I.~M., \& {Crawford}, C.~S. 1991, \mnras, 252, 482

\bibitem[{{Aharonian} {et~al.}(2008){Aharonian}, {Akhperjanian}, {Barres de
  Almeida}, {Bazer-Bachi}, {Behera}, {Beilicke}, {Benbow}, {Bernl{\"o}hr},
  {Boisson}, {Bolz}, {Borrel}, {Braun}, {Brion}, {Brown}, {B{\"u}hler},
  {Bulik}, {B{\"u}sching}, {Boutelier}, {Carrigan}, {Chadwick}, {Chounet},
  {Clapson}, {Coignet}, {Cornils}, {Costamante}, {Dalton}, {Degrange},
  {Dickinson}, {Djannati-Ata{\"\i}}, {Domainko}, {O'C. Drury}, {Dubois},
  {Dubus}, {Dyks}, {Egberts}, {Emmanoulopoulos}, {Espigat}, {Farnier},
  {Feinstein}, {Fiasson}, {F{\"o}rster}, {Fontaine}, {Funk}, {F{\"u}{\ss}ling},
  {Gallant}, {Giebels}, {Glicenstein}, {Gl{\"u}ck}, {Goret}, {Hadjichristidis},
  {Hauser}, {Hauser}, {Heinzelmann}, {Henri}, {Hermann}, {Hinton}, {Hoffmann},
  {Hofmann}, {Holleran}, {Hoppe}, {Horns}, {Jacholkowska}, {de Jager}, {Jung},
  {Katarzy{\'n}ski}, {Kendziorra}, {Kerschhaggl}, {Kh{\'e}lifi}, {Keogh},
  {Komin}, {Kosack}, {Lamanna}, {Latham}, {Lemi{\`e}re}, {Lemoine-Goumard},
  {Lenain}, {Lohse}, {Martin}, {Martineau-Huynh}, {Marcowith}, {Masterson},
  {Maurin}, {Maurin}, {McComb}, {Moderski}, {Moulin}, {de Naurois}, {Nedbal},
  {Nolan}, {Ohm}, {Olive}, {de O{\~n}a Wilhelmi}, {Orford}, {Osborne},
  {Ostrowski}, {Panter}, {Pedaletti}, {Pelletier}, {Petrucci}, {Pita},
  {P{\"u}hlhofer}, {Punch}, {Ranchon}, {Raubenheimer}, {Raue}, {Rayner},
  {Renaud}, {Ripken}, {Rob}, {Rolland}, {Rosier-Lees}, {Rowell}, {Rudak},
  {Ruppel}, {Sahakian}, {Santangelo}, {Schlickeiser}, {Sch{\"o}ck},
  {Schr{\"o}der}, {Schwanke}, {Schwarzburg}, {Schwemmer}, {Shalchi}, {Sol},
  {Spangler}, {Stawarz}, {Steenkamp}, {Stegmann}, {Superina}, {Tam},
  {Tavernet}, {Terrier}, {van Eldik}, {Vasileiadis}, {Venter}, {Vialle},
  {Vincent}, {Vivier}, {V{\"o}lk}, {Volpe}, {Wagner}, {Ward}, {Zdziarski}, \&
  {Zech}}]{hessVLT1553}
{Aharonian}, F., {Akhperjanian}, A.~G., {Barres de Almeida}, U., {et~al.} 2008,
  \aap, 477, 481

\bibitem[{{Ahumada} {et~al.}(2020){Ahumada}, {Allende Prieto}, {Almeida},
  {Anders}, {Anderson}, {Andrews}, {Anguiano}, {Arcodia}, {Armengaud},
  {Aubert}, {Avila}, {Avila-Reese}, {Badenes}, {Balland}, {Barger},
  {Barrera-Ballesteros}, {Basu}, {Bautista}, {Beaton}, {Beers}, {Benavides},
  {Bender}, {Bernardi}, {Bershady}, {Beutler}, {Bidin}, {Bird}, {Bizyaev},
  {Blanc}, {Blanton}, {Boquien}, {Borissova}, {Bovy}, {Brandt}, {Brinkmann},
  {Brownstein}, {Bundy}, {Bureau}, {Burgasser}, {Burtin}, {Cano-D{\'\i}az},
  {Capasso}, {Cappellari}, {Carrera}, {Chabanier}, {Chaplin}, {Chapman},
  {Cherinka}, {Chiappini}, {Doohyun Choi}, {Chojnowski}, {Chung}, {Clerc},
  {Coffey}, {Comerford}, {Comparat}, {da Costa}, {Cousinou}, {Covey}, {Crane},
  {Cunha}, {Ilha}, {Dai}, {Damsted}, {Darling}, {Davidson}, {Davies}, {Dawson},
  {De}, {de la Macorra}, {De Lee}, {Queiroz}, {Deconto Machado}, {de la Torre},
  {Dell'Agli}, {du Mas des Bourboux}, {Diamond-Stanic}, {Dillon}, {Donor},
  {Drory}, {Duckworth}, {Dwelly}, {Ebelke}, {Eftekharzadeh}, {Davis Eigenbrot},
  {Elsworth}, {Eracleous}, {Erfanianfar}, {Escoffier}, {Fan}, {Farr},
  {Fern{\'a}ndez-Trincado}, {Feuillet}, {Finoguenov}, {Fofie},
  {Fraser-McKelvie}, {Frinchaboy}, {Fromenteau}, {Fu}, {Galbany}, {Garcia},
  {Garc{\'\i}a-Hern{\'a}ndez}, {Garma Oehmichen}, {Ge}, {Geimba Maia},
  {Geisler}, {Gelfand}, {Goddy}, {Gonzalez-Perez}, {Grabowski}, {Green},
  {Grier}, {Guo}, {Guy}, {Harding}, {Hasselquist}, {Hawken}, {Hayes}, {Hearty},
  {Hekker}, {Hogg}, {Holtzman}, {Horta}, {Hou}, {Hsieh}, {Huber}, {Hunt}, {Ider
  Chitham}, {Imig}, {Jaber}, {Jimenez Angel}, {Johnson}, {Jones},
  {J{\"o}nsson}, {Jullo}, {Kim}, {Kinemuchi}, {Kirkpatrick}, {Kite}, {Klaene},
  {Kneib}, {Kollmeier}, {Kong}, {Kounkel}, {Krishnarao}, {Lacerna}, {Lan},
  {Lane}, {Law}, {Le Goff}, {Leung}, {Lewis}, {Li}, {Lian}, {Lin}, {Long},
  {Longa-Pe{\~n}a}, {Lundgren}, {Lyke}, {Mackereth}, {MacLeod}, {Majewski},
  {Manchado}, {Maraston}, {Martini}, {Masseron}, {Masters}, {Mathur},
  {McDermid}, {Merloni}, {Merrifield}, {M{\'e}sz{\'a}ros}, {Miglio}, {Minniti},
  {Minsley}, {Miyaji}, {Mohammad}, {Mosser}, {Mueller}, {Muna},
  {Mu{\~n}oz-Guti{\'e}rrez}, {Myers}, {Nadathur}, {Nair}, {Nandra}, {Correa do
  Nascimento}, {Nevin}, {Newman}, {Nidever}, {Nitschelm}, {Noterdaeme},
  {O'Connell}, {Olmstead}, {Oravetz}, {Oravetz}, {Osorio}, {Pace}, {Padilla},
  {Palanque-Delabrouille}, {Palicio}, {Pan}, {Pan}, {Parker}, {Paviot},
  {Peirani}, {Ram{\'r}ez}, {Penny}, {Percival}, {Perez-Fournon},
  {P{\'e}rez-R{\`a}fols}, {Petitjean}, {Pieri}, {Pinsonneault}, {Poovelil},
  {Povick}, {Prakash}, {Price-Whelan}, {Raddick}, {Raichoor}, {Ray}, {Rembold},
  {Rezaie}, {Riffel}, {Riffel}, {Rix}, {Robin}, {Roman-Lopes},
  {Rom{\'a}n-Z{\'u}{\~n}iga}, {Rose}, {Ross}, {Rossi}, {Rowlands}, {Rubin},
  {Salvato}, {S{\'a}nchez}, {S{\'a}nchez-Menguiano}, {S{\'a}nchez-Gallego},
  {Sayres}, {Schaefer}, {Schiavon}, {Schimoia}, {Schlafly}, {Schlegel},
  {Schneider}, {Schultheis}, {Schwope}, {Seo}, {Serenelli}, {Shafieloo},
  {Shamsi}, {Shao}, {Shen}, {Shetrone}, {Shirley}, {Silva Aguirre}, {Simon},
  {Skrutskie}, {Slosar}, {Smethurst}, {Sobeck}, {Sodi}, {Souto}, {Stark},
  {Stassun}, {Steinmetz}, {Stello}, {Stermer}, {Storchi-Bergmann},
  {Streblyanska}, {Stringfellow}, {Stutz}, {Su{\'a}rez}, {Sun},
  {Taghizadeh-Popp}, {Talbot}, {Tayar}, {Thakar}, {Theriault}, {Thomas},
  {Thomas}, {Tinker}, {Tojeiro}, {Toledo}, {Tremonti}, {Troup}, {Tuttle},
  {Unda-Sanzana}, {Valentini}, {Vargas-Gonz{\'a}lez}, {Vargas-Maga{\~n}a},
  {V{\'a}zquez-Mata}, {Vivek}, {Wake}, {Wang}, {Weaver}, {Weijmans}, {Wild},
  {Wilson}, {Wilson}, {Wolthuis}, {Wood-Vasey}, {Yan}, {Yang}, {Y{\`e}che},
  {Zamora}, {Zarrouk}, {Zasowski}, {Zhang}, {Zhao}, {Zhao}, {Zheng}, {Zheng},
  {Zhu}, \& {Zou}}]{Ahu19}
{Ahumada}, R., {Allende Prieto}, C., {Almeida}, A., {et~al.} 2020, \apjs, 249,
  3

\bibitem[{{Ajello} {et~al.}(2020){Ajello}, {Angioni}, {Axelsson}, {Ballet},
  {Barbiellini}, {Bastieri}, {Becerra Gonzalez}, {Bellazzini}, {Bissaldi},
  {Bloom}, {Bonino}, {Bottacini}, {Bruel}, {Buson}, {Cafardo}, {Cameron},
  {Cavazzuti}, {Chen}, {Cheung}, {Ciprini}, {Costantin}, {Cutini}, {D'Ammando},
  {de la Torre Luque}, {de Menezes}, {de Palma}, {Desai}, {Di Lalla}, {Di
  Venere}, {Dom{\'\i}nguez}, {Dirirsa}, {Ferrara}, {Finke}, {Franckowiak},
  {Fukazawa}, {Funk}, {Fusco}, {Gargano}, {Garrappa}, {Gasparrini},
  {Giglietto}, {Giordano}, {Giroletti}, {Green}, {Grenier}, {Guiriec},
  {Harita}, {Hays}, {Horan}, {Itoh}, {J{\'o}hannesson}, {Kovac'evic'},
  {Krauss}, {Kreter}, {Kuss}, {Larsson}, {Leto}, {Li}, {Liodakis}, {Longo},
  {Loparco}, {Lott}, {Lovellette}, {Lubrano}, {Madejski}, {Maldera},
  {Manfreda}, {Mart{\'\i}-Devesa}, {Massaro}, {Mazziotta}, {Mereu}, {Meyer},
  {Migliori}, {Mirabal}, {Mizuno}, {Monzani}, {Morselli}, {Moskalenko},
  {Negro}, {Nemmen}, {Nuss}, {Ojha}, {Ojha}, {Omodei}, {Orienti}, {Orlando},
  {Ormes}, {Paliya}, {Pei}, {Pe{\~n}a-Herazo}, {Persic}, {Pesce-Rollins},
  {Petrov}, {Piron}, {Poon}, {Principe}, {Rain{\`o}}, {Rando}, {Rani},
  {Razzano}, {Razzaque}, {Reimer}, {Reimer}, {Schinzel}, {Serini}, {Sgr{\`o}},
  {Siskind}, {Spandre}, {Spinelli}, {Suson}, {Tachibana}, {Thompson}, {Torres},
  {Torresi}, {Troja}, {Valverde}, {van Zyl}, \&
  {Yassine}}]{2020ApJ...892..105A}
{Ajello}, M., {Angioni}, R., {Axelsson}, M., {et~al.} 2020, \apj, 892, 105

\bibitem[{{Ajello} {et~al.}(2017){Ajello}, {Atwood}, {Baldini}, {Ballet},
  {Barbiellini}, {Bastieri}, {Bellazzini}, {Bissaldi}, {Blandford}, {Bloom},
  {Bonino}, {Bregeon}, {Britto}, {Bruel}, {Buehler}, {Buson}, {Cameron},
  {Caputo}, {Caragiulo}, {Caraveo}, {Cavazzuti}, {Cecchi}, {Charles},
  {Chekhtman}, {Cheung}, {Chiaro}, {Ciprini}, {Cohen}, {Costantin}, {Costanza},
  {Cuoco}, {Cutini}, {D'Ammando}, {de Palma}, {Desiante}, {Digel}, {Di Lalla},
  {Di Mauro}, {Di Venere}, {Dom{\'\i}nguez}, {Drell}, {Dumora}, {Favuzzi},
  {Fegan}, {Ferrara}, {Fortin}, {Franckowiak}, {Fukazawa}, {Funk}, {Fusco},
  {Gargano}, {Gasparrini}, {Giglietto}, {Giommi}, {Giordano}, {Giroletti},
  {Glanzman}, {Green}, {Grenier}, {Grondin}, {Grove}, {Guillemot}, {Guiriec},
  {Harding}, {Hays}, {Hewitt}, {Horan}, {J{\'o}hannesson}, {Kensei}, {Kuss},
  {La Mura}, {Larsson}, {Latronico}, {Lemoine-Goumard}, {Li}, {Longo},
  {Loparco}, {Lott}, {Lubrano}, {Magill}, {Maldera}, {Manfreda}, {Mazziotta},
  {McEnery}, {Meyer}, {Michelson}, {Mirabal}, {Mitthumsiri}, {Mizuno},
  {Moiseev}, {Monzani}, {Morselli}, {Moskalenko}, {Negro}, {Nuss}, {Ohsugi},
  {Omodei}, {Orienti}, {Orlando}, {Palatiello}, {Paliya}, {Paneque}, {Perkins},
  {Persic}, {Pesce-Rollins}, {Piron}, {Porter}, {Principe}, {Rain{\`o}},
  {Rando}, {Razzano}, {Razzaque}, {Reimer}, {Reimer}, {Reposeur}, {Saz
  Parkinson}, {Sgr{\`o}}, {Simone}, {Siskind}, {Spada}, {Spandre}, {Spinelli},
  {Stawarz}, {Suson}, {Takahashi}, {Tak}, {Thayer}, {Thayer}, {Thompson},
  {Torres}, {Torresi}, {Troja}, {Vianello}, {Wood}, \&
  {Wood}}]{2017ApJS..232...18A}
{Ajello}, M., {Atwood}, W.~B., {Baldini}, L., {et~al.} 2017, \apjs, 232, 18

\bibitem[{{Bertin} \& {Arnouts}(1996)}]{1996A&AS..117..393B}
{Bertin}, E. \& {Arnouts}, S. 1996, \aaps, 117, 393

\bibitem[{{Chambers} {et~al.}(2016){Chambers}, {Magnier}, {Metcalfe},
  {Flewelling}, {Huber}, {Waters}, {Denneau}, {Draper}, {Farrow}, {Finkbeiner},
  {Holmberg}, {Koppenhoefer}, {Price}, {Rest}, {Saglia}, {Schlafly}, {Smartt},
  {Sweeney}, {Wainscoat}, {Burgett}, {Chastel}, {Grav}, {Heasley}, {Hodapp},
  {Jedicke}, {Kaiser}, {Kudritzki}, {Luppino}, {Lupton}, {Monet}, {Morgan},
  {Onaka}, {Shiao}, {Stubbs}, {Tonry}, {White}, {Ba{\~n}ados}, {Bell},
  {Bender}, {Bernard}, {Boegner}, {Boffi}, {Botticella}, {Calamida},
  {Casertano}, {Chen}, {Chen}, {Cole}, {Deacon}, {Frenk}, {Fitzsimmons},
  {Gezari}, {Gibbs}, {Goessl}, {Goggia}, {Gourgue}, {Goldman}, {Grant},
  {Grebel}, {Hambly}, {Hasinger}, {Heavens}, {Heckman}, {Henderson}, {Henning},
  {Holman}, {Hopp}, {Ip}, {Isani}, {Jackson}, {Keyes}, {Koekemoer}, {Kotak},
  {Le}, {Liska}, {Long}, {Lucey}, {Liu}, {Martin}, {Masci}, {McLean}, {Mindel},
  {Misra}, {Morganson}, {Murphy}, {Obaika}, {Narayan}, {Nieto-Santisteban},
  {Norberg}, {Peacock}, {Pier}, {Postman}, {Primak}, {Rae}, {Rai}, {Riess},
  {Riffeser}, {Rix}, {R{\"o}ser}, {Russel}, {Rutz}, {Schilbach}, {Schultz},
  {Scolnic}, {Strolger}, {Szalay}, {Seitz}, {Small}, {Smith}, {Soderblom},
  {Taylor}, {Thomson}, {Taylor}, {Thakar}, {Thiel}, {Thilker}, {Unger},
  {Urata}, {Valenti}, {Wagner}, {Walder}, {Walter}, {Watters}, {Werner},
  {Wood-Vasey}, \& {Wyse}}]{2016arXiv161205560C}
{Chambers}, K.~C., {Magnier}, E.~A., {Metcalfe}, N., {et~al.} 2016, arXiv
  e-prints, arXiv:1612.05560

\bibitem[{{Cherenkov Telescope Array Consortium} {et~al.}(2019){Cherenkov
  Telescope Array Consortium}, {Acharya}, {Agudo}, {Al Samarai}, {Alfaro},
  {Alfaro}, {Alispach}, {Alves Batista}, {Amans}, {Amato}, {Ambrosi},
  {Antolini}, {Antonelli}, {Aramo}, {Araya}, {Armstrong}, {Arqueros},
  {Arrabito}, {Asano}, {Ashley}, {Backes}, {Balazs}, {Balbo}, {Ballester},
  {Ballet}, {Bamba}, {Barkov}, {Barres de Almeida}, {Barrio}, {Bastieri},
  {Becherini}, {Belfiore}, {Benbow}, {Berge}, {Bernardini}, {Bernardini},
  {Bernardos}, {Bernl{\"o}hr}, {Bertucci}, {Biasuzzi}, {Bigongiari}, {Biland},
  {Bissaldi}, {Biteau}, {Blanch}, {Blazek}, {Boisson}, {Bolmont}, {Bonanno},
  {Bonardi}, {Bonavolont{\`a}}, {Bonnoli}, {Bosnjak}, {B{\"o}ttcher},
  {Braiding}, {Bregeon}, {Brill}, {Brown}, {Brun}, {Brunetti}, {Buanes},
  {Buckley}, {Bugaev}, {B{\"u}hler}, {Bulgarelli}, {Bulik}, {Burton},
  {Burtovoi}, {Busetto}, {Canestrari}, {Capalbi}, {Capitanio}, {Caproni},
  {Caraveo}, {C{\'a}rdenas}, {Carlile}, {Carosi}, {Carqu{\'\i}n}, {Carr},
  {Casanova}, {Cascone}, {Catalani}, {Catalano}, {Cauz}, {Cerruti}, {Chadwick},
  {Chaty}, {Chaves}, {Chen}, {Chen}, {Chernyakova}, {Chikawa}, {Christov},
  {Chudoba}, {Cie{\'s}lar}, {Coco}, {Colafrancesco}, {Colin}, {Conforti},
  {Connaughton}, {Conrad}, {Contreras}, {Cortina}, {Costa}, {Costantini},
  {Cotter}, {Covino}, {Crocker}, {Cuadra}, {Cuevas}, {Cumani}, {D'A{\`\i}},
  {D'Ammando}, {D'Avanzo}, {D'Urso}, {Daniel}, {Davids}, {Dawson}, {Dazzi}, {De
  Angelis}, {de C{\'a}ssia dos Anjos}, {De Cesare}, {De Franco}, {de Gouveia
  Dal Pino}, {de la Calle}, {de los Reyes Lopez}, {De Lotto}, {De Luca}, {De
  Lucia}, {de Naurois}, {de O{\~n}a Wilhelmi}, {De Palma}, {De Persio}, {de
  Souza}, {Deil}, {Del Santo}, {Delgado}, {della Volpe}, {Di Girolamo}, {Di
  Pierro}, {Di Venere}, {D{\'\i}az}, {Dib}, {Diebold}, {Djannati-Ata{\"\i}},
  {Dom{\'\i}nguez}, {Dominis Prester}, {Dorner}, {Doro}, {Drass}, {Dravins},
  {Dubus}, {Dwarkadas}, {Ebr}, {Eckner}, {Egberts}, {Einecke}, {Ekoume},
  {Els{\"a}sser}, {Ernenwein}, {Espinoza}, {Evoli}, {Fairbairn},
  {Falceta-Goncalves}, {Falcone}, {Farnier}, {Fasola}, {Fedorova}, {Fegan},
  {Fernandez-Alonso}, {Fern{\'a}ndez-Barral}, {Ferrand}, {Fesquet},
  {Filipovic}, {Fioretti}, {Fontaine}, {Fornasa}, {Fortson}, {Freixas
  Coromina}, {Fruck}, {Fujita}, {Fukazawa}, {Funk}, {F{\"u}{\ss}ling},
  {Gabici}, {Gadola}, {Gallant}, {Garcia}, {Garcia L{\'o}pez}, {Garczarczyk},
  {Gaskins}, {Gasparetto}, {Gaug}, {Gerard}, {Giavitto}, {Giglietto}, {Giommi},
  {Giordano}, {Giro}, {Giroletti}, {Giuliani}, {Glicenstein}, {Gnatyk},
  {Godinovic}, {Goldoni}, {G{\'o}mez-Vargas}, {Gonz{\'a}lez}, {Gonz{\'a}lez},
  {G{\"o}tz}, {Graham}, {Grandi}, {Granot}, {Green}, {Greenshaw}, {Griffiths},
  {Gunji}, {Hadasch}, {Hara}, {Hardcastle}, {Hassan}, {Hayashi}, {Hayashida},
  {Heller}, {Helo}, {Hermann}, {Hinton}, {Hnatyk}, {Hofmann}, {Holder},
  {Horan}, {H{\"o}randel}, {Horns}, {Horvath}, {Hovatta}, {Hrabovsky},
  {Hrupec}, {Humensky}, {H{\"u}tten}, {Iarlori}, {Inada}, {Inome}, {Inoue},
  {Inoue}, {Inoue}, {Iocco}, {Ioka}, {Iori}, {Ishio}, {Iwamura}, {Jamrozy},
  {Janecek}, {Jankowsky}, {Jean}, {Jung-Richardt}, {Jurysek}, {Kaaret},
  {Karkar}, {Katagiri}, {Katz}, {Kawanaka}, {Kazanas}, {Kh{\'e}lifi}, {Kieda},
  {Kimeswenger}, {Kimura}, {Kisaka}, {Knapp}, {Kn{\"o}dlseder}, {Koch},
  {Kohri}, {Komin}, {Kosack}, {Kraus}, {Krause}, {Krau{\ss}}, {Kubo}, {Kukec
  Mezek}, {Kuroda}, {Kushida}, {La Palombara}, {Lamanna}, {Lang}, {Lapington},
  {Le Blanc}, {Leach}, {Lees}, {Lefaucheur}, {Leigui de Oliveira}, {Lenain},
  {Lico}, {Limon}, {Lindfors}, {Lohse}, {Lombardi}, {Longo}, {L{\'o}pez},
  {L{\'o}pez-Coto}, {Lu}, {Lucarelli}, {Luque-Escamilla}, {Lyard}, {Maccarone},
  {Maier}, {Majumdar}, {Malaguti}, {Mandat}, {Maneva}, {Manganaro}, {Mangano},
  {Marcowith}, {Mar{\'\i}n}, {Markoff}, {Mart{\'\i}}, {Martin},
  {Mart{\'\i}nez}, {Mart{\'\i}nez}, {Masetti}, {Masuda}, {Maurin}, {Maxted},
  {Mazin}, {Medina}, {Melandri}, {Mereghetti}, {Meyer}, {Minaya}, {Mirabal},
  {Mirzoyan}, {Mitchell}, {Mizuno}, {Moderski}, {Mohammed}, {Mohrmann},
  {Montaruli}, {Moralejo}, {Morcuende-Parrilla}, {Mori}, {Morlino}, {Morris},
  {Morselli}, {Moulin}, {Mukherjee}, {Mundell}, {Murach}, {Muraishi}, {Murase},
  {Nagai}, {Nagataki}, {Nagayoshi}, {Naito}, {Nakamori}, {Nakamura}, {Niemiec},
  {Nieto}, {Niko{\l}ajuk}, {Nishijima}, {Noda}, {Nosek}, {Novosyadlyj},
  {Nozaki}, {O'Brien}, {Oakes}, {Ohira}, {Ohishi}, {Ohm}, {Okazaki}, {Okumura},
  {Ong}, {Orienti}, {Orito}, {Osborne}, {Ostrowski}, {Otte}, {Oya}, {Padovani},
  {Paizis}, {Palatiello}, {Palatka}, {Paoletti}, {Paredes}, {Pareschi},
  {Parsons}, {Pe'er}, {Pech}, {Pedaletti}, {Perri}, {Persic}, {Petrashyk},
  {Petrucci}, {Petruk}, {Peyaud}, {Pfeifer}, {Piano}, {Pisarski}, {Pita},
  {Pohl}, {Polo}, {Pozo}, {Prandini}, {Prast}, {Principe}, {Prokhorov},
  {Prokoph}, {Prouza}, {P{\"u}hlhofer}, {Punch}, {P{\"u}rckhauer}, {Queiroz},
  {Quirrenbach}, {Rain{\`o}}, {Razzaque}, {Reimer}, {Reimer}, {Reisenegger},
  {Renaud}, {Rezaeian}, {Rhode}, {Ribeiro}, {Rib{\'o}}, {Richtler}, {Rico},
  {Rieger}, {Riquelme}, {Rivoire}, {Rizi}, {Rodriguez}, {Rodriguez Fernandez},
  {Rodr{\'\i}guez V{\'a}zquez}, {Rojas}, {Romano}, {Romeo}, {Rosado}, {Rovero},
  {Rowell}, {Rudak}, {Rugliancich}, {Rulten}, {Sadeh}, {Safi-Harb}, {Saito},
  {Sakaki}, {Sakurai}, {Salina}, {S{\'a}nchez-Conde}, {Sandaker}, {Sandoval},
  {Sangiorgi}, {Sanguillon}, {Sano}, {Santander}, {Sarkar}, {Satalecka},
  {Saturni}, {Schioppa}, {Schlenstedt}, {Schneider}, {Schoorlemmer},
  {Schovanek}, {Schulz}, {Schussler}, {Schwanke}, {Sciacca}, {Scuderi},
  {Seitenzahl}, {Semikoz}, {Sergijenko}, {Servillat}, {Shalchi}, {Shellard},
  {Sidoli}, {Siejkowski}, {Sillanp{\"a}{\"a}}, {Sironi}, {Sitarek}, {Sliusar},
  {Slowikowska}, {Sol}, {Stamerra}, {Stani{\v{c}}}, {Starling}, {Stawarz},
  {Stefanik}, {Stephan}, {Stolarczyk}, {Stratta}, {Straumann}, {Suomijarvi},
  {Supanitsky}, {Tagliaferri}, {Tajima}, {Tavani}, {Tavecchio}, {Tavernet},
  {Tayabaly}, {Tejedor}, {Temnikov}, {Terada}, {Terrier}, {Terzic}, {Teshima},
  {Testa}, {Thoudam}, {Tian}, {Tibaldo}, {Tluczykont}, {Todero Peixoto},
  {Tokanai}, {Tomastik}, {Tonev}, {Tornikoski}, {Torres}, {Torresi}, {Tosti},
  {Tothill}, {Tovmassian}, {Travnicek}, {Trichard}, {Trifoglio}, {Troyano
  Pujadas}, {Tsujimoto}, {Umana}, {Vagelli}, {Vagnetti}, {Valentino},
  {Vallania}, {Valore}, {van Eldik}, {Vandenbroucke}, {Varner}, {Vasileiadis},
  {Vassiliev}, {V{\'a}zquez Acosta}, {Vecchi}, {Vega}, {Vercellone}, {Veres},
  {Vergani}, {Verzi}, {Vettolani}, {Viana}, {Vigorito}, {Villanueva}, {Voelk},
  {Vollhardt}, {Vorobiov}, {Vrastil}, {Vuillaume}, {Wagner}, {Wagner},
  {Walter}, {Ward}, {Warren}, {Watson}, {Werner}, {White}, {White},
  {Wierzcholska}, {Wilcox}, {Will}, {Williams}, {Wischnewski}, {Wood},
  {Yamamoto}, {Yamazaki}, {Yanagita}, {Yang}, {Yoshida}, {Yoshiike},
  {Yoshikoshi}, {Zacharias}, {Zaharijas}, {Zampieri}, {Zandanel}, {Zanin},
  {Zavrtanik}, {Zavrtanik}, {Zdziarski}, {Zech}, {Zechlin}, {Zhdanov},
  {Ziegler}, \& {Zorn}}]{2019scta.book.....C}
{Cherenkov Telescope Array Consortium}, {Acharya}, B.~S., {Agudo}, I., {et~al.}
  2019, {Science with the Cherenkov Telescope Array} (World Scientific
  Publishing Co. Pte. Ltd.)

\bibitem[{{D'Ammando} {et~al.}(2024){D'Ammando}, {Goldoni}, {Max-Moerbeck},
  {Becerra Gonzalez}, {Kasai}, {Williams}, {Alvarez-Crespo}, {Backes}, {Barres
  de Almeida}, {Boisson}, {Cotter}, {Fallah Ramazani}, {Hervet}, {Lindfors},
  {Mukhi-Nilo}, {Pita}, {Splettstoesser}, \& {van
  Soelen}}]{2024arXiv240107911D}
{D'Ammando}, F., {Goldoni}, P., {Max-Moerbeck}, W., {et~al.} 2024, arXiv
  e-prints, arXiv:2401.07911

\bibitem[{{Danforth} {et~al.}(2010){Danforth}, {Keeney}, {Stocke}, {Shull}, \&
  {Yao}}]{Dan10}
{Danforth}, C.~W., {Keeney}, B.~A., {Stocke}, J.~T., {Shull}, J.~M., \& {Yao},
  Y. 2010, \apj, 720, 976

\bibitem[{{Deshpande} {et~al.}(2017){Deshpande}, {Hughes}, \&
  {Wittman}}]{Desh17}
{Deshpande}, A.~J., {Hughes}, J.~P., \& {Wittman}, D. 2017, \apj, 839, 124

\bibitem[{{Dom{\'\i}nguez} {et~al.}(2011){Dom{\'\i}nguez}, {Primack},
  {Rosario}, {Prada}, {Gilmore}, {Faber}, {Koo}, {Somerville},
  {P{\'e}rez-Torres}, {P{\'e}rez-Gonz{\'a}lez}, {Huang}, {Davis},
  {Guhathakurta}, {Barmby}, {Conselice}, {Lozano}, {Newman}, \&
  {Cooper}}]{Dom11}
{Dom{\'\i}nguez}, A., {Primack}, J.~R., {Rosario}, D.~J., {et~al.} 2011,
  \mnras, 410, 2556

\bibitem[{{Dorigo Jones} {et~al.}(2022){Dorigo Jones}, {Johnson}, {Muzahid},
  {Charlton}, {Chen}, {Narayanan}, {Sameer}, {Schaye}, \& {Wijers}}]{Dor22}
{Dorigo Jones}, J., {Johnson}, S.~D., {Muzahid}, S., {et~al.} 2022, \mnras,
  509, 4330

\bibitem[{{Fioc} \& {Rocca-Volmerange}(2019)}]{2019A&A...623A.143F}
{Fioc}, M. \& {Rocca-Volmerange}, B. 2019, \aap, 623, A143

\bibitem[{{Fukugita} {et~al.}(1995){Fukugita}, {Shimasaku}, \&
  {Ichikawa}}]{Fuku95}
{Fukugita}, M., {Shimasaku}, K., \& {Ichikawa}, T. 1995, \pasp, 107, 945

\bibitem[{{Ghisellini} {et~al.}(2011){Ghisellini}, {Tavecchio}, {Foschini}, \&
  {Ghirland a}}]{Ghis11}
{Ghisellini}, G., {Tavecchio}, F., {Foschini}, L., \& {Ghirland a}, G. 2011,
  \mnras, 414, 2674

\bibitem[{{Goldoni} {et~al.}(2021){Goldoni}, {Pita}, {Boisson}, {Max-Moerbeck},
  {Kasai}, {Williams}, {D'Ammando}, {Navarro-Aranguiz}, {Backes}, {Barres de
  Almeida}, {Becerra-Gonzalez}, {Cotter}, {Hervet}, {Lenain}, {Lindfors},
  {Sol}, \& {Wagner}}]{Gol21}
{Goldoni}, P., {Pita}, S., {Boisson}, C., {et~al.} 2021, \aap, 650, A106

\bibitem[{{Graham}(2007)}]{Gra07}
{Graham}, A.~W. 2007, \mnras, 379, 711

\bibitem[{{IceCube Collaboration} {et~al.}(2018){IceCube Collaboration},
  {Aartsen}, {Ackermann}, {Adams}, {Aguilar}, {Ahlers}, {Ahrens}, {Al Samarai},
  {Altmann}, {Andeen}, {Anderson}, {Ansseau}, {Anton}, {Arg{\"u}elles},
  {Auffenberg}, {Axani}, {Bagherpour}, {Bai}, {Barron}, {Barwick}, {Baum},
  {Bay}, {Beatty}, {Becker Tjus}, {Becker}, {BenZvi}, {Berley}, {Bernardini},
  {Besson}, {Binder}, {Bindig}, {Blaufuss}, {Blot}, {Bohm}, {B{\"o}rner},
  {Bos}, {B{\"o}ser}, {Botner}, {Bourbeau}, {Bourbeau}, {Bradascio}, {Braun},
  {Brenzke}, {Bretz}, {Bron}, {Brostean-Kaiser}, {Burgman}, {Busse}, {Carver},
  {Cheung}, {Chirkin}, {Christov}, {Clark}, {Classen}, {Coenders}, {Collin},
  {Conrad}, {Coppin}, {Correa}, {Cowen}, {Cross}, {Dave}, {Day}, {de
  Andr{\'e}}, {De Clercq}, {DeLaunay}, {Dembinski}, {De Ridder}, {Desiati}, {de
  Vries}, {de Wasseige}, {de With}, {DeYoung}, {D{\'\i}az-V{\'e}lez}, {di
  Lorenzo}, {Dujmovic}, {Dumm}, {Dunkman}, {Dvorak}, {Eberhardt}, {Ehrhardt},
  {Eichmann}, {Eller}, {Evenson}, {Fahey}, {Fazely}, {Felde}, {Filimonov},
  {Finley}, {Flis}, {Franckowiak}, {Friedman}, {Fritz}, {Gaisser}, {Gallagher},
  {Gerhardt}, {Ghorbani}, {Glauch}, {Gl{\"u}senkamp}, {Goldschmidt},
  {Gonzalez}, {Grant}, {Griffith}, {Haack}, {Hallgren}, {Halzen}, {Hanson},
  {Hebecker}, {Heereman}, {Helbing}, {Hellauer}, {Hickford}, {Hignight},
  {Hill}, {Hoffman}, {Hoffmann}, {Hoinka}, {Hokanson-Fasig}, {Hoshina},
  {Huang}, {Huber}, {Hultqvist}, {H{\"u}nnefeld}, {Hussain}, {In}, {Iovine},
  {Ishihara}, {Jacobi}, {Japaridze}, {Jeong}, {Jero}, {Jones}, {Kalaczynski},
  {Kang}, {Kappes}, {Kappesser}, {Karg}, {Karle}, {Katz}, {Kauer}, {Keivani},
  {Kelley}, {Kheirandish}, {Kim}, {Kim}, {Kintscher}, {Kiryluk}, {Kittler},
  {Klein}, {Koirala}, {Kolanoski}, {K{\"o}pke}, {Kopper}, {Kopper},
  {Koschinsky}, {Koskinen}, {Kowalski}, {Krings}, {Kroll}, {Kr{\"u}ckl},
  {Kunwar}, {Kurahashi}, {Kuwabara}, {Kyriacou}, {Labare}, {Lanfranchi},
  {Larson}, {Lauber}, {Leonard}, {Lesiak-Bzdak}, {Leuermann}, {Liu}, {Lozano
  Mariscal}, {Lu}, {L{\"u}nemann}, {Luszczak}, {Madsen}, {Maggi}, {Mahn},
  {Mancina}, {Maruyama}, {Mase}, {Maunu}, {Meagher}, {Medici}, {Meier},
  {Menne}, {Merino}, {Meures}, {Miarecki}, {Micallef}, {Moment{\'e}},
  {Montaruli}, {Moore}, {Morse}, {Moulai}, {Nahnhauer}, {Nakarmi}, {Naumann},
  {Neer}, {Niederhausen}, {Nowicki}, {Nygren}, {Obertacke Pollmann}, {Olivas},
  {O'Murchadha}, {O'Sullivan}, {Palczewski}, {Pandya}, {Pankova}, {Peiffer},
  {Pepper}, {P{\'e}rez de los Heros}, {Pieloth}, {Pinat}, {Plum}, {Price},
  {Przybylski}, {Raab}, {R{\"a}del}, {Rameez}, {Rauch}, {Rawlins}, {Rea},
  {Reimann}, {Relethford}, {Relich}, {Resconi}, {Rhode}, {Richman},
  {Robertson}, {Rongen}, {Rott}, {Ruhe}, {Ryckbosch}, {Rysewyk}, {Safa},
  {S{\"a}lzer}, {Sanchez Herrera}, {Sandrock}, {Sandroos}, {Santander},
  {Sarkar}, {Sarkar}, {Satalecka}, {Schlunder}, {Schmidt}, {Schneider},
  {Schoenen}, {Sch{\"o}neberg}, {Schumacher}, {Sclafani}, {Seckel},
  {Seunarine}, {Soedingrekso}, {Soldin}, {Song}, {Spiczak}, {Spiering},
  {Stachurska}, {Stamatikos}, {Stanev}, {Stasik}, {Stein}, {Stettner},
  {Steuer}, {Stezelberger}, {Stokstad}, {St{\"o}{\ss}l}, {Strotjohann},
  {Stuttard}, {Sullivan}, {Sutherland}, {Taboada}, {Tatar}, {Tenholt},
  {Ter-Antonyan}, {Terliuk}, {Tilav}, {Toale}, {Tobin}, {Toennis}, {Toscano},
  {Tosi}, {Tselengidou}, {Tung}, {Turcati}, {Turley}, {Ty}, {Unger}, {Usner},
  {Vandenbroucke}, {Van Driessche}, {van Eijk}, {van Eijndhoven}, {Vanheule},
  {van Santen}, {Vogel}, {Vraeghe}, {Walck}, {Wallace}, {Wallraff}, {Wandler},
  {Wandkowsky}, {Waza}, {Weaver}, {Weiss}, {Wendt}, {Werthebach}, {Westerhoff},
  {Whelan}, {Whitehorn}, {Wiebe}, {Wiebusch}, {Wille}, {Williams}, {Wills},
  {Wolf}, {Wood}, {Wood}, {Woschnagg}, {Xu}, {Xu}, {Xu}, {Yanez}, {Yodh},
  {Yoshida}, {Yuan}, {Fermi-LAT Collaboration}, {Abdollahi}, {Ajello},
  {Angioni}, {Baldini}, {Ballet}, {Barbiellini}, {Bastieri}, {Bechtol},
  {Bellazzini}, {Berenji}, {Bissaldi}, {Blandford}, {Bonino}, {Bottacini},
  {Bregeon}, {Bruel}, {Buehler}, {Burnett}, {Burns}, {Buson}, {Cameron},
  {Caputo}, {Caraveo}, {Cavazzuti}, {Charles}, {Chen}, {Cheung}, {Chiang},
  {Chiaro}, {Ciprini}, {Cohen-Tanugi}, {Conrad}, {Costantin}, {Cutini},
  {D'Ammando}, {de Palma}, {Digel}, {Di Lalla}, {Di Mauro}, {Di Venere},
  {Dom{\'\i}nguez}, {Favuzzi}, {Franckowiak}, {Fukazawa}, {Funk}, {Fusco},
  {Gargano}, {Gasparrini}, {Giglietto}, {Giomi}, {Giommi}, {Giordano},
  {Giroletti}, {Glanzman}, {Green}, {Grenier}, {Grondin}, {Guiriec}, {Harding},
  {Hayashida}, {Hays}, {Hewitt}, {Horan}, {J{\'o}hannesson}, {Kadler},
  {Kensei}, {Kocevski}, {Krauss}, {Kreter}, {Kuss}, {La Mura}, {Larsson},
  {Latronico}, {Lemoine-Goumard}, {Li}, {Longo}, {Loparco}, {Lovellette},
  {Lubrano}, {Magill}, {Maldera}, {Malyshev}, {Manfreda}, {Mazziotta},
  {McEnery}, {Meyer}, {Michelson}, {Mizuno}, {Monzani}, {Morselli},
  {Moskalenko}, {Negro}, {Nuss}, {Ojha}, {Omodei}, {Orienti}, {Orlando},
  {Palatiello}, {Paliya}, {Perkins}, {Persic}, {Pesce-Rollins}, {Piron},
  {Porter}, {Principe}, {Rain{\`o}}, {Rando}, {Rani}, {Razzano}, {Razzaque},
  {Reimer}, {Reimer}, {Renault-Tinacci}, {Ritz}, {Rochester}, {Saz Parkinson},
  {Sgr{\`o}}, {Siskind}, {Spandre}, {Spinelli}, {Suson}, {Tajima}, {Takahashi},
  {Tanaka}, {Thayer}, {Thompson}, {Tibaldo}, {Torres}, {Torresi}, {Tosti},
  {Troja}, {Valverde}, {Vianello}, {Vogel}, {Wood}, {Wood}, {Zaharijas}, {MAGIC
  Collaboration}, {Ahnen}, {Ansoldi}, {Antonelli}, {Arcaro}, {Baack},
  {Babi{\'c}}, {Banerjee}, {Bangale}, {Barres de Almeida}, {Barrio}, {Becerra
  Gonz{\'a}lez}, {Bednarek}, {Bernardini}, {Berti}, {Bhattacharyya}, {Biland},
  {Blanch}, {Bonnoli}, {Carosi}, {Carosi}, {Ceribella}, {Chatterjee}, {Colak},
  {Colin}, {Colombo}, {Contreras}, {Cortina}, {Covino}, {Cumani}, {Da Vela},
  {Dazzi}, {De Angelis}, {De Lotto}, {Delfino}, {Delgado}, {Di Pierro},
  {Dom{\'\i}nguez}, {Dominis Prester}, {Dorner}, {Doro}, {Einecke},
  {Elsaesser}, {Fallah Ramazani}, {Fern{\'a}ndez-Barral}, {Fidalgo}, {Foffano},
  {Pfrang}, {Fonseca}, {Font}, {Franceschini}, {Fruck}, {Galindo}, {Gallozzi},
  {Garc{\'\i}a L{\'o}pez}, {Garczarczyk}, {Gaug}, {Giammaria}, {Godinovi{\'c}},
  {Gora}, {Guberman}, {Hadasch}, {Hahn}, {Hassan}, {Hayashida}, {Herrera},
  {Hose}, {Hrupec}, {Inoue}, {Ishio}, {Konno}, {Kubo}, {Kushida}, {Lelas},
  {Lindfors}, {Lombardi}, {Longo}, {L{\'o}pez}, {Maggio}, {Majumdar},
  {Makariev}, {Maneva}, {Manganaro}, {Mannheim}, {Maraschi}, {Mariotti},
  {Mart{\'\i}nez}, {Masuda}, {Mazin}, {Minev}, {M}, {Mirzoyan}, {Moralejo},
  {Moreno}, {Moretti}, {Nagayoshi}, {Neustroev}, {Niedzwiecki}, {Nievas
  Rosillo}, {Nigro}, {Nilsson}, {Ninci}, {Nishijima}, {Noda}, {Nogu{\'e}s},
  {Paiano}, {Palacio}, {Paneque}, {Paoletti}, {Paredes}, {Pedaletti},
  {Peresano}, {Persic}, {Prada Moroni}, {Prandini}, {Puljak}, {Rodriguez
  Garcia}, {Reichardt}, {Rhode}, {Rib{\'o}}, {Rico}, {Righi}, {Rugliancich},
  {Saito}, {Satalecka}, {Schweizer}, {Sitarek}, {{\v{S}}nidaric
  {\textasciiacute}}, {Sobczynska}, {Stamerra}, {Strzys}, {Suri{\'c}},
  {Takahashi}, {Tavecchio}, {Temnikov}, {Terzi{\'c}}, {Teshima},
  {Torres-Alb{\`a}}, {Treves}, {Tsujimoto}, {Vanzo}, {Vazquez Acosta}, {Vovk},
  {Ward}, {Will}, {S}, {Zaric {\textasciiacute}}, {AGILE Team}, {Lucarelli},
  {Tavani}, {Piano}, {Donnarumma}, {Pittori}, {Verrecchia}, {Barbiellini},
  {Bulgarelli}, {Caraveo}, {Cattaneo}, {Colafrancesco}, {Costa}, {Di Cocco},
  {Ferrari}, {Gianotti}, {Giuliani}, {Lipari}, {Mereghetti}, {Morselli},
  {Pacciani}, {Paoletti}, {Parmiggiani}, {Pellizzoni}, {Picozza}, {Pilia},
  {Rappoldi}, {Trois}, {Vercellone}, {Vittorini}, {ASAS-SN Team}, {Stanek},
  {Franckowiak}, {Kochanek}, {Beacom}, {Thompson}, {Holoien}, {Dong}, {Prieto},
  {Shappee}, {Holmbo}, {HAWC Collaboration}, {Abeysekara}, {Albert}, {Alfaro},
  {Alvarez}, {Arceo}, {Arteaga-Vel{\'a}zquez}, {Avila Rojas}, {Ayala Solares},
  {Becerril}, {Belmont-Moreno}, {Bernal}, {Caballero-Mora}, {Capistr{\'a}n},
  {Carrami{\~n}ana}, {Casanova}, {Castillo}, {Cotti}, {Cotzomi}, {Couti{\~n}o
  de Le{\'o}n}, {De Le{\'o}n}, {De la Fuente}, {Diaz Hernandez}, {Dichiara},
  {Dingus}, {DuVernois}, {D{\'\i}az-V{\'e}lez}, {Ellsworth}, {Engel},
  {Fiorino}, {Fleischhack}, {Fraija}, {Garc{\'\i}a-Gonz{\'a}lez}, {Garfias},
  {Gonz{\'a}lez Mu{\~n}oz}, {Gonz{\'a}lez}, {Goodman}, {Hampel-Arias},
  {Harding}, {Hernandez}, {Hona}, {Hueyotl-Zahuantitla}, {Hui},
  {H{\"u}ntemeyer}, {Iriarte}, {Jardin-Blicq}, {Joshi}, {Kaufmann}, {Kunde},
  {Lara}, {Lauer}, {Lee}, {Lennarz}, {Le{\'o}n Vargas}, {Linnemann},
  {Longinotti}, {Luis-Raya}, {Luna-Garc{\'\i}a}, {Malone}, {Marinelli},
  {Martinez}, {Martinez-Castellanos}, {Mart{\'\i}nez-Castro},
  {Mart{\'\i}nez-Huerta}, {Matthews}, {Miranda-Romagnoli}, {Moreno},
  {Mostaf{\'a}}, {Nayerhoda}, {Nellen}, {Newbold}, {Nisa}, {Noriega-Papaqui},
  {Pelayo}, {Pretz}, {P{\'e}rez-P{\'e}rez}, {Ren}, {Rho}, {Rivi{\`e}re},
  {Rosa-Gonz{\'a}lez}, {Rosenberg}, {Ruiz-Velasco}, {Ruiz-Velasco}, {Salesa
  Greus}, {Sandoval}, {Schneider}, {Schoorlemmer}, {Sinnis}, {Smith},
  {Springer}, {Surajbali}, {Tibolla}, {Tollefson}, {Torres}, {Villase{\~n}or},
  {Weisgarber}, {Werner}, {Yapici}, {Gaurang}, {Zepeda}, {Zhou}, {{\'A}lvarez},
  {H.~E.~S.~S. Collaboration}, {Abdalla}, {Ang{\"u}ner}, {Armand}, {Backes},
  {Becherini}, {Berge}, {B{\"o}ttcher}, {Boisson}, {Bolmont}, {Bonnefoy},
  {Bordas}, {Brun}, {B{\"u}chele}, {Bulik}, {Caroff}, {Carosi}, {Casanova},
  {Cerruti}, {Chakraborty}, {Chandra}, {Chen}, {Colafrancesco}, {Davids},
  {Deil}, {Devin}, {Djannati-Ata{\"\i}}, {Egberts}, {Emery}, {Eschbach},
  {Fiasson}, {Fontaine}, {Funk}, {F{\"u}{\ss}ling}, {Gallant}, {Gat{\'e}},
  {Giavitto}, {Glawion}, {Glicenstein}, {Gottschall}, {Grondin}, {Haupt},
  {Henri}, {Hinton}, {Hoischen}, {Holch}, {Huber}, {Jamrozy}, {Jankowsky},
  {Jankowsky}, {Jouvin}, {Jung-Richardt}, {Kerszberg}, {Kh{\'e}lifi}, {King},
  {Klepser}, {Kluz {\textasciiacute}niak}, {Komin}, {Kraus}, {Lefaucheur},
  {Lemi{\`e}re}, {Lemoine-Goumard}, {Lenain}, {Leser}, {Lohse},
  {L{\'o}pez-Coto}, {Lorentz}, {Lypova}, {Marandon}, {Guillem
  Mart{\'\i}-Devesa}, {Maurin}, {Mitchell}, {Moderski}, {Mohamed}, {Mohrmann},
  {Moulin}, {Murach}, {de Naurois}, {Niederwanger}, {Niemiec}, {Oakes},
  {O'Brien}, {Ohm}, {Ostrowski}, {Oya}, {Panter}, {Parsons}, {Perennes},
  {Piel}, {Pita}, {Poireau}, {Priyana Noel}, {Prokoph}, {P{\"u}hlhofer},
  {Quirrenbach}, {Raab}, {Rauth}, {Renaud}, {Rieger}, {Rinchiuso}, {Romoli},
  {Rowell}, {Rudak}, {Sasaki}, {Sanchez}, {Schlickeiser}, {Sch{\"u}ssler},
  {Schulz}, {Schwanke}, {Seglar-Arroyo}, {Shafi}, {Simoni}, {Sol}, {Stegmann},
  {Steppa}, {Tavernier}, {Taylor}, {Tiziani}, {Trichard}, {Tsirou}, {van
  Eldik}, {van Rensburg}, {van Soelen}, {Veh}, {Vincent}, {Voisin}, {Wagner},
  {Wagner}, {Wierzcholska}, {Zanin}, {Zdziarski}, {Zech}, {Ziegler}, {Zorn},
  {{\.Z}ywucka}, {INTEGRAL Team}, {Savchenko}, {Ferrigno}, {Bazzano}, {Diehl},
  {Kuulkers}, {Laurent}, {Mereghetti}, {Natalucci}, {Panessa}, {Rodi},
  {Ubertini}, {Kanata}, Teams, {Morokuma}, {Ohta}, {Tanaka}, {Mori},
  {Yamanaka}, {Kawabata}, {Utsumi}, {Nakaoka}, {Kawabata}, {Nagashima},
  {Yoshida}, {Matsuoka}, {Itoh}, {Kapteyn Team}, {Keel}, {Liverpool Telescope
  Team}, {Copperwheat}, {Steele}, {Swift/NuSTAR Team}, {Cenko}, {Cowen},
  {DeLaunay}, {Evans}, {Fox}, {Keivani}, {Kennea}, {Marshall}, {Osborne},
  {Santander}, {Tohuvavohu}, {Turley}, {VERITAS Collaboration}, {Abeysekara},
  {Archer}, {Benbow}, {Bird}, {Brill}, {Brose}, {Buchovecky}, {Buckley},
  {Bugaev}, {Christiansen}, {Connolly}, {Cui}, {Daniel}, {Errando}, {Falcone},
  {Feng}, {Finley}, {Fortson}, {Furniss}, {Gueta}, {H{\"u}tten}, {Hervet},
  {Hughes}, {Humensky}, {Johnson}, {Kaaret}, {Kar}, {Kelley-Hoskins},
  {Kertzman}, {Kieda}, {Krause}, {Krennrich}, {Kumar}, {Lang}, {Lin}, {Maier},
  {McArthur}, {Moriarty}, {Mukherjee}, {Nieto}, {O'Brien}, {Ong}, {Otte},
  {Park}, {Petrashyk}, {Pohl}, {Popkow}, {Pueschel}, {Quinn}, {Ragan},
  {Reynolds}, {Richards}, {Roache}, {Rulten}, {Sadeh}, {Santander}, {Scott},
  {Sembroski}, {Shahinyan}, {Sushch}, {Tr{\'e}panier}, {Tyler}, {Vassiliev},
  {Wakely}, {Weinstein}, {Wells}, {Wilcox}, {Wilhelm}, {Williams}, {Zitzer},
  {VLA/B Team}, {Tetarenko}, {Kimball}, {Miller-Jones}, \&
  {Sivakoff}}]{2018Sci...361.1378I}
{IceCube Collaboration}, {Aartsen}, M.~G., {Ackermann}, M., {et~al.} 2018,
  Science, 361, eaat1378

\bibitem[{{Jiang} {et~al.}(2011){Jiang}, {Greene}, \& {Ho}}]{Jia11}
{Jiang}, Y.-F., {Greene}, J.~E., \& {Ho}, L.~C. 2011, \apjl, 737, L45

\bibitem[{{Johnson} {et~al.}(2019){Johnson}, {Mulchaey}, {Chen}, {Wijers},
  {Connor}, {Muzahid}, {Schaye}, {Cen}, {Carlsten}, {Charlton}, {Drout},
  {Goulding}, {Hansen}, \& {Walth}}]{John19}
{Johnson}, S.~D., {Mulchaey}, J.~S., {Chen}, H.-W., {et~al.} 2019, \apjl, 884,
  L31

\bibitem[{{Kasai} {et~al.}(2023){Kasai}, {Goldoni}, {Pita}, {Williams},
  {Max-Moerbeck}, {Hervet}, {Cotter}, {Backes}, {Boisson}, {Becerra
  Gonz{\'a}lez}, {Barres de Almeida}, {D'Ammando}, {Fallah Ramazani}, \&
  {Lindfors}}]{Kas23}
{Kasai}, E., {Goldoni}, P., {Pita}, S., {et~al.} 2023, \mnras, 518, 2675

\bibitem[{{La Mura} {et~al.}(2022){La Mura}, {Becerra Gonzalez}, {Chiaro},
  {Ciroi}, \& {Otero-Santos}}]{2022MNRAS.515.4810L}
{La Mura}, G., {Becerra Gonzalez}, J., {Chiaro}, G., {Ciroi}, S., \&
  {Otero-Santos}, J. 2022, \mnras, 515, 4810

\bibitem[{{Landoni} {et~al.}(2014){Landoni}, {Falomo}, {Treves}, \&
  {Sbarufatti}}]{Lan14}
{Landoni}, M., {Falomo}, R., {Treves}, A., \& {Sbarufatti}, B. 2014, \aap, 570,
  A126

\bibitem[{{Mannucci} {et~al.}(2001){Mannucci}, {Basile}, {Poggianti},
  {Cimatti}, {Daddi}, {Pozzetti}, \& {Vanzi}}]{Mannu01}
{Mannucci}, F., {Basile}, F., {Poggianti}, B.~M., {et~al.} 2001, \mnras, 326,
  745

\bibitem[{{Masetti} {et~al.}(2013){Masetti}, {Sbarufatti}, {Parisi},
  {Jim{\'e}nez-Bail{\'o}n}, {Chavushyan}, {Vogt}, {Sguera}, {Stephen},
  {Palazzi}, {Bassani}, {Bazzano}, {Fiocchi}, {Galaz}, {Landi}, {Malizia},
  {Minniti}, {Morelli}, \& {Ubertini}}]{Mase13}
{Masetti}, N., {Sbarufatti}, B., {Parisi}, P., {et~al.} 2013, \aap, 559, A58

\bibitem[{{Massaro} {et~al.}(2015{\natexlab{a}}){Massaro}, {Maselli}, {Leto},
  {Marchegiani}, {Perri}, {Giommi}, \& {Piranomonte}}]{2015Ap&SS.357...75M}
{Massaro}, E., {Maselli}, A., {Leto}, C., {et~al.} 2015{\natexlab{a}}, \apss,
  357, 75

\bibitem[{{Massaro} {et~al.}(2019){Massaro}, {{\'A}lvarez-Crespo}, {Capetti},
  {Baldi}, {Pillitteri}, {Campana}, \& {Paggi}}]{Massaro19a}
{Massaro}, F., {{\'A}lvarez-Crespo}, N., {Capetti}, A., {et~al.} 2019, \apjs,
  240, 20

\bibitem[{{Massaro} {et~al.}(2020{\natexlab{a}}){Massaro}, {Capetti}, {Paggi},
  {Baldi}, {Tramacere}, {Pillitteri}, \& {Campana}}]{Massaro20}
{Massaro}, F., {Capetti}, A., {Paggi}, A., {et~al.} 2020{\natexlab{a}}, \apjl,
  900, L34

\bibitem[{{Massaro} {et~al.}(2020{\natexlab{b}}){Massaro}, {Capetti}, {Paggi},
  {Baldi}, {Tramacere}, {Pillitteri}, {Campana}, {Jimenez-Gallardo}, \&
  {Missaglia}}]{Massaro19b}
{Massaro}, F., {Capetti}, A., {Paggi}, A., {et~al.} 2020{\natexlab{b}}, \apjs,
  247, 71

\bibitem[{{Massaro} {et~al.}(2015{\natexlab{b}}){Massaro}, {Landoni},
  {D'Abrusco}, {Milisavljevic}, {Paggi}, {Masetti}, {Smith}, \&
  {Tosti}}]{Mas15}
{Massaro}, F., {Landoni}, M., {D'Abrusco}, R., {et~al.} 2015{\natexlab{b}},
  \aap, 575, A124

\bibitem[{{Meisner} \& {Romani}(2010)}]{Mei10}
{Meisner}, A.~M. \& {Romani}, R.~W. 2010, \apj, 712, 14

\bibitem[{{Mishra} {et~al.}(2018){Mishra}, {Chand}, {Krishna}, {Joshi},
  {Shchekinov}, \& {Fatkhullin}}]{2018MNRAS.473.5154M}
{Mishra}, S., {Chand}, H., {Krishna}, G., {et~al.} 2018, \mnras, 473, 5154

\bibitem[{{Muriel} {et~al.}(2015){Muriel}, {Donzelli}, {Rovero}, \&
  {Pichel}}]{Mur15}
{Muriel}, H., {Donzelli}, C., {Rovero}, A.~C., \& {Pichel}, A. 2015, \aap, 574,
  A101

\bibitem[{{Nilsson} {et~al.}(2003){Nilsson}, {Pursimo}, {Heidt}, {Takalo},
  {Sillanp{\"a}{\"a}}, \& {Brinkmann}}]{Nil03}
{Nilsson}, K., {Pursimo}, T., {Heidt}, J., {et~al.} 2003, \aap, 400, 95

\bibitem[{{Nilsson} {et~al.}(2012){Nilsson}, {Pursimo}, {Villforth},
  {Lindfors}, {Takalo}, \& {Sillanp{\"a}{\"a}}}]{2012A&A...547A...1N}
{Nilsson}, K., {Pursimo}, T., {Villforth}, C., {et~al.} 2012, \aap, 547, A1

\bibitem[{{Paiano} {et~al.}(2023){Paiano}, {Falomo}, {Treves}, {Padovani},
  {Giommi}, {Scarpa}, {Bisogni}, \& {Marini}}]{Pai23}
{Paiano}, S., {Falomo}, R., {Treves}, A., {et~al.} 2023, \mnras, 521, 2270

\bibitem[{{Paiano} {et~al.}(2020){Paiano}, {Falomo}, {Treves}, \&
  {Scarpa}}]{Pai20}
{Paiano}, S., {Falomo}, R., {Treves}, A., \& {Scarpa}, R. 2020, \mnras, 497, 94

\bibitem[{{Paiano} {et~al.}(2017){Paiano}, {Landoni}, {Falomo}, {Treves},
  {Scarpa}, \& {Righi}}]{Pai17}
{Paiano}, S., {Landoni}, M., {Falomo}, R., {et~al.} 2017, \apj, 837, 144

\bibitem[{{Pita} {et~al.}(2014){Pita}, {Goldoni}, {Boisson}, {Lenain}, {Punch},
  {G{\'e}rard}, {Hammer}, {Kaper}, \& {Sol}}]{2014A&A...565A..12P}
{Pita}, S., {Goldoni}, P., {Boisson}, C., {et~al.} 2014, \aap, 565, A12

\bibitem[{{Rosa Gonz{\'a}lez} {et~al.}(2019){Rosa Gonz{\'a}lez}, {Muriel},
  {Mayya}, {Aretxaga}, {Becerra Gonz{\'a}lez}, {Carrami{\~n}ana},
  {M{\'e}ndez-Abreu}, {Vega}, {Terlevich}, {Couti{\~n}o de Le{\'o}n},
  {Furniss}, {Longinotti}, {Terlevich}, {Pichel}, {Rovero}, \&
  {Donzelli}}]{Ros19}
{Rosa Gonz{\'a}lez}, D., {Muriel}, H., {Mayya}, Y.~D., {et~al.} 2019, \mnras,
  482, 5422

\bibitem[{{Rovero} {et~al.}(2016){Rovero}, {Muriel}, {Donzelli}, \&
  {Pichel}}]{Rov16}
{Rovero}, A.~C., {Muriel}, H., {Donzelli}, C., \& {Pichel}, A. 2016, \aap, 589,
  A92

\bibitem[{{Samir} {et~al.}(2020){Samir}, {Takey}, \&
  {Shaker}}]{2020Ap&SS.365..142S}
{Samir}, R.~M., {Takey}, A., \& {Shaker}, A.~A. 2020, \apss, 365, 142

\bibitem[{{Sbarufatti} {et~al.}(2005{\natexlab{a}}){Sbarufatti}, {Treves}, \&
  {Falomo}}]{Sbar05}
{Sbarufatti}, B., {Treves}, A., \& {Falomo}, R. 2005{\natexlab{a}}, \apj, 635,
  173

\bibitem[{{Sbarufatti} {et~al.}(2005{\natexlab{b}}){Sbarufatti}, {Treves},
  {Falomo}, {Heidt}, {Kotilainen}, \& {Scarpa}}]{2005AJ....129..559S}
{Sbarufatti}, B., {Treves}, A., {Falomo}, R., {et~al.} 2005{\natexlab{b}}, \aj,
  129, 559

\bibitem[{{Sbarufatti} {et~al.}(2006){Sbarufatti}, {Treves}, {Falomo}, {Heidt},
  {Kotilainen}, \& {Scarpa}}]{Sba06}
{Sbarufatti}, B., {Treves}, A., {Falomo}, R., {et~al.} 2006, \aj, 132, 1

\bibitem[{{Scarpa} {et~al.}(2000){Scarpa}, {Urry}, {Falomo}, {Pesce}, \&
  {Treves}}]{Scarpa00}
{Scarpa}, R., {Urry}, C.~M., {Falomo}, R., {Pesce}, J.~E., \& {Treves}, A.
  2000, \apj, 532, 740

\bibitem[{{Schlafly} \& {Finkbeiner}(2011)}]{2011ApJ...737..103S}
{Schlafly}, E.~F. \& {Finkbeiner}, D.~P. 2011, \apj, 737, 103

\bibitem[{{Shaw} {et~al.}(2012){Shaw}, {Romani}, {Cotter}, {Healey},
  {Michelson}, {Readhead}, {Richards}, {Max-Moerbeck}, {King}, \&
  {Potter}}]{2012ApJ...748...49S}
{Shaw}, M.~S., {Romani}, R.~W., {Cotter}, G., {et~al.} 2012, \apj, 748, 49

\bibitem[{{Shaw} {et~al.}(2013){Shaw}, {Romani}, {Cotter}, {Healey},
  {Michelson}, {Readhead}, {Richards}, {Max-Moerbeck}, {King}, \&
  {Potter}}]{Shaw13}
{Shaw}, M.~S., {Romani}, R.~W., {Cotter}, G., {et~al.} 2013, \apj, 764, 135

\bibitem[{{Stickel} {et~al.}(1991){Stickel}, {Padovani}, {Urry}, {Fried}, \&
  {Kuehr}}]{Stick91}
{Stickel}, M., {Padovani}, P., {Urry}, C.~M., {Fried}, J.~W., \& {Kuehr}, H.
  1991, \apj, 374, 431

\bibitem[{{Treves} {et~al.}(2007){Treves}, {Falomo}, \& {Uslenghi}}]{Treves07}
{Treves}, A., {Falomo}, R., \& {Uslenghi}, M. 2007, \aap, 473, L17

\bibitem[{{Tsarevsky} {et~al.}(2005){Tsarevsky}, {de Freitas Pacheco},
  {Kardashev}, {de Laverny}, {Th{\'e}venin}, {Slee}, {Stathakis}, {Barsukova},
  {Goransky}, \& {Komberg}}]{Tsa05}
{Tsarevsky}, G., {de Freitas Pacheco}, J.~A., {Kardashev}, N., {et~al.} 2005,
  \aap, 438, 949

\bibitem[{{Urry} \& {Padovani}(1995)}]{1995PASP..107..803U}
{Urry}, C.~M. \& {Padovani}, P. 1995, \pasp, 107, 803

\bibitem[{{Wakely} \& {Horan}(2008)}]{2008ICRC....3.1341W}
{Wakely}, S.~P. \& {Horan}, D. 2008, in International Cosmic Ray Conference,
  Vol.~3, International Cosmic Ray Conference, 1341--1344

\bibitem[{{White} {et~al.}(2000){White}, {Becker}, {Gregg},
  {Laurent-Muehleisen}, {Brotherton}, {Impey}, {Petry}, {Foltz}, {Chaffee},
  {Richards}, {Oegerle}, {Helfand}, {McMahon}, \& {Cabanela}}]{White00}
{White}, R.~L., {Becker}, R.~H., {Gregg}, M.~D., {et~al.} 2000, \apjs, 126, 133

\bibitem[{{Wittman} {et~al.}(2006){Wittman}, {Dell'Antonio}, {Hughes},
  {Margoniner}, {Tyson}, {Cohen}, \& {Norman}}]{Witt06}
{Wittman}, D., {Dell'Antonio}, I.~P., {Hughes}, J.~P., {et~al.} 2006, \apj,
  643, 128

\bibitem[{{Wurtz} {et~al.}(1997){Wurtz}, {Stocke}, {Ellingson}, \&
  {Yee}}]{1997ApJ...480..547W}
{Wurtz}, R., {Stocke}, J.~T., {Ellingson}, E., \& {Yee}, H.~K.~C. 1997, \apj,
  480, 547

\end{thebibliography}

\begin{appendix}
\section{Light curves of the sources}
As discussed in Section~5, we obtained light curves from the ZTF for the 12 sources that were observed with NOT in ZTF era to evaluate if the imaging observations were performed in a low state of the source. We have also noted the timing of any spectroscopy done by \citet{Kas23}, \citet{2024arXiv240107911D}, or any high signal-to-noise spectroscopy by other authors.
   
   \begin{figure*}
   \centering
   \includegraphics[width=0.9\textwidth]{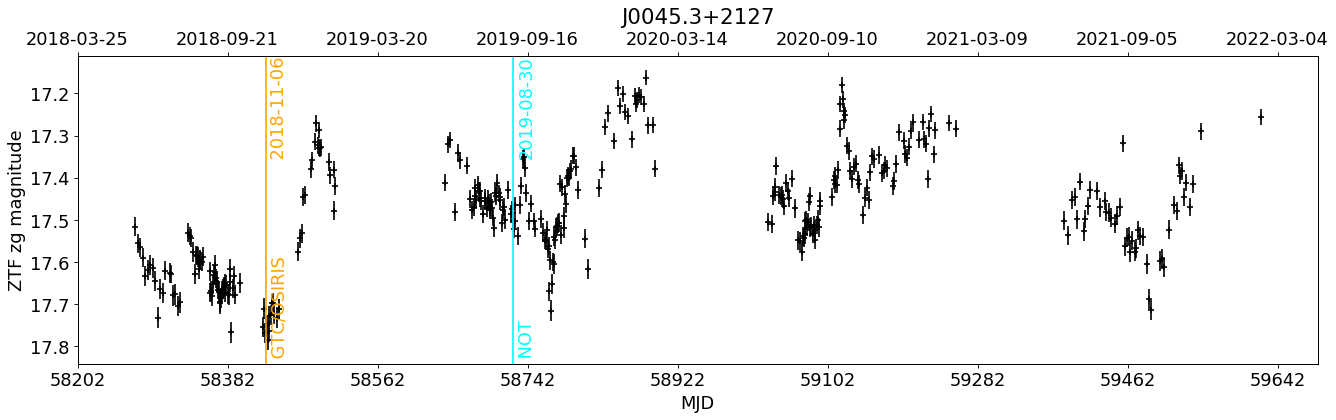}
   \includegraphics[width=0.9\textwidth]{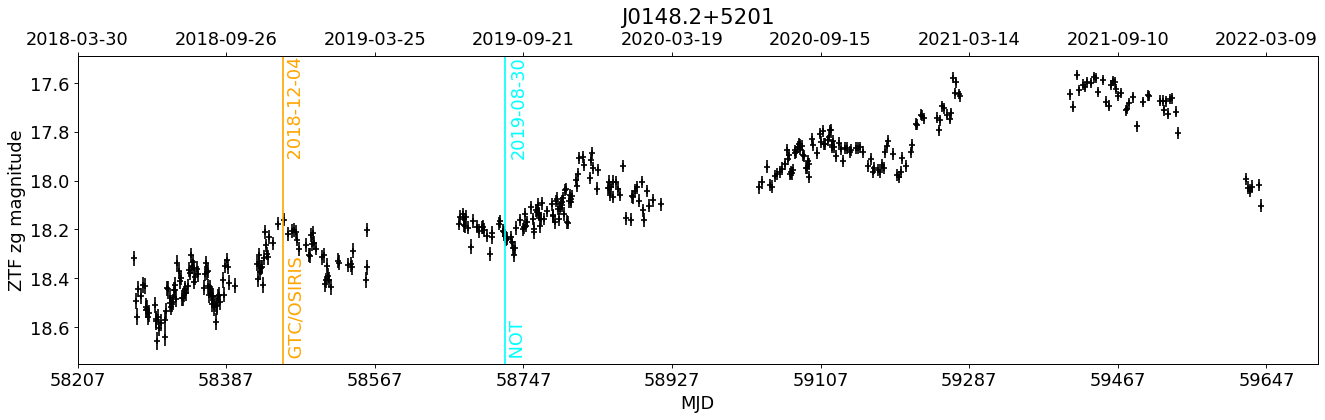}
   \includegraphics[width=0.9\textwidth]{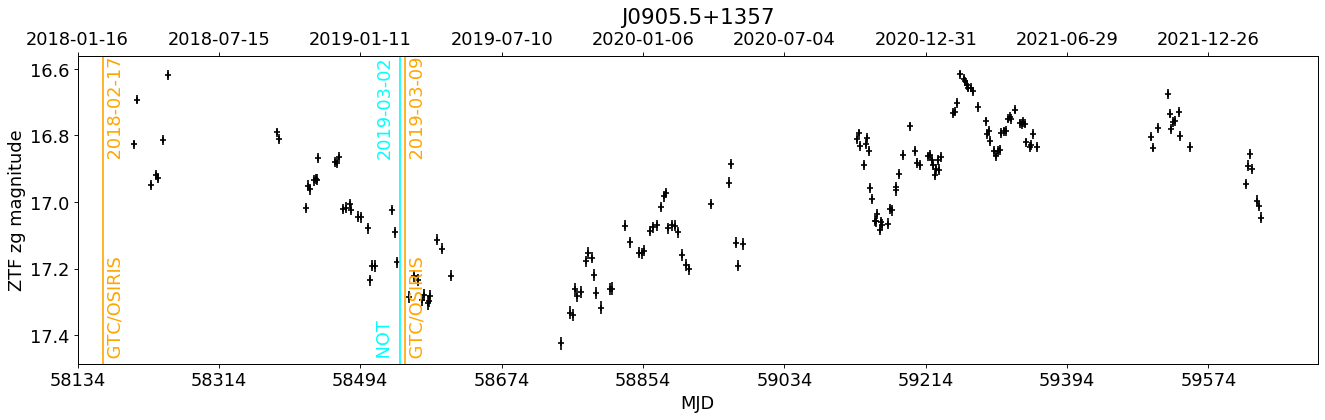}
   \includegraphics[width=0.9\textwidth]{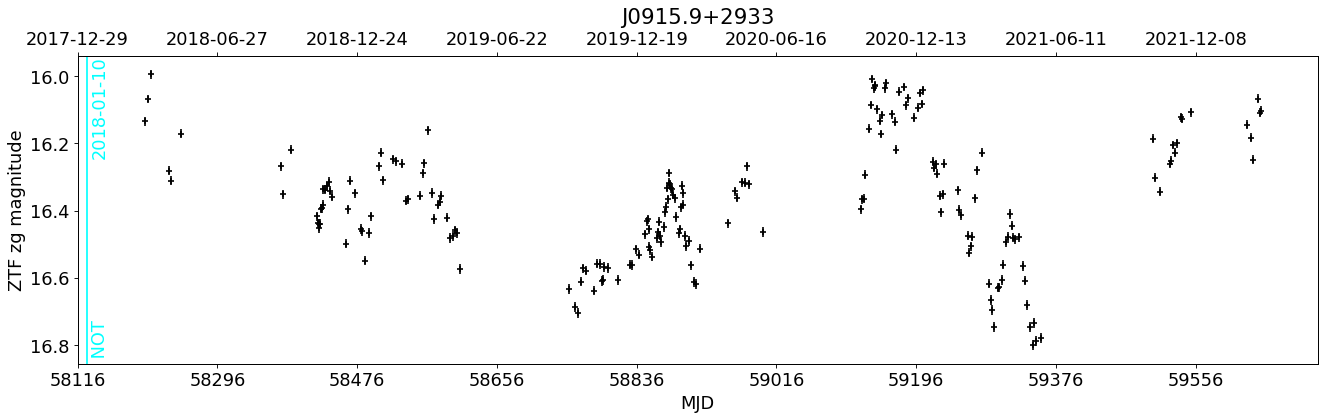}
   \caption{ZTF g-band light curves of J0045.3+2127 (top), J0148.2+5201, J0905.5+1357 and J0915.9+2933 (bottom). The vertical lines show the timing of spectroscopy (orange) and our deep imaging observations (magenta).}
              \label{LC1}
    \end{figure*}

   \begin{figure*}
   \centering
      \includegraphics[width=0.9\textwidth]{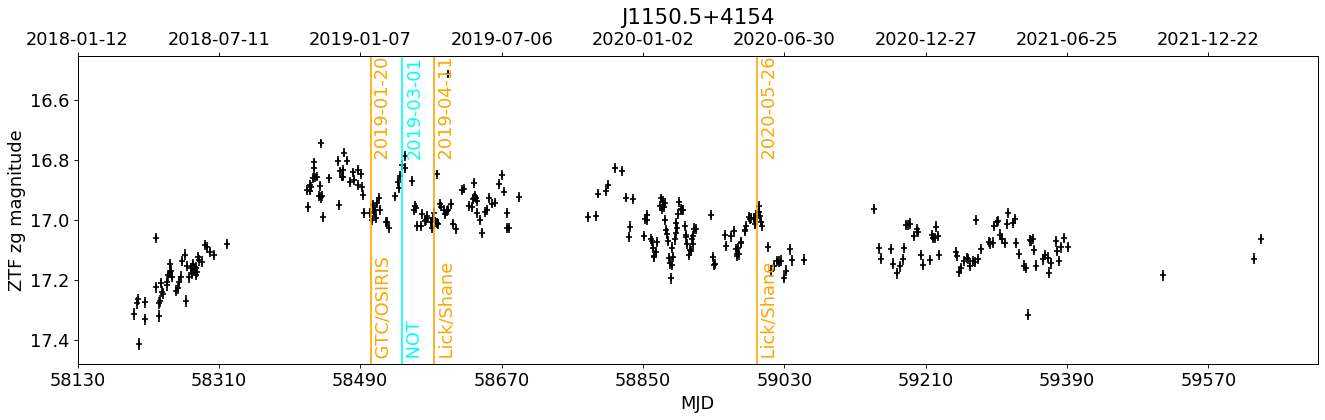}
   \includegraphics[width=0.9\textwidth]{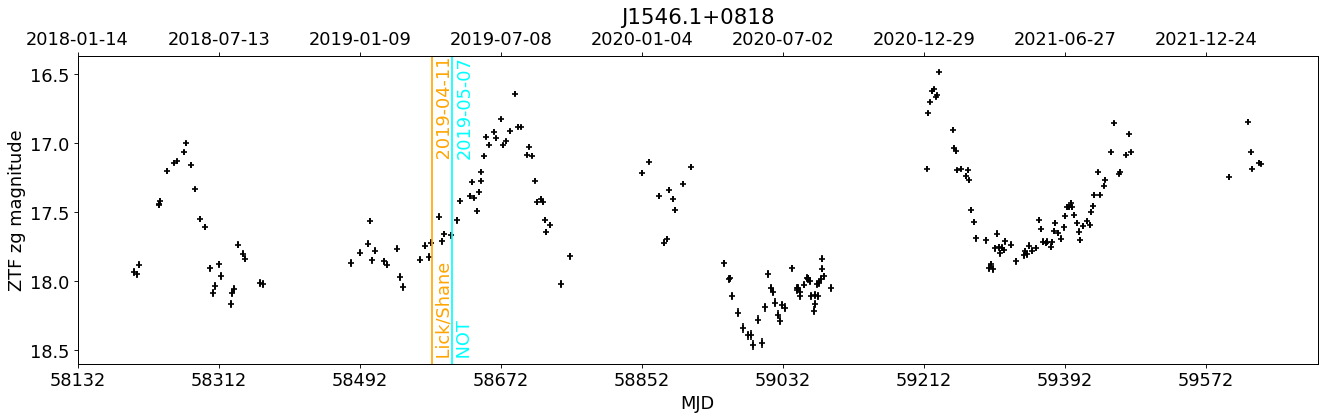}
   \includegraphics[width=0.9\textwidth]{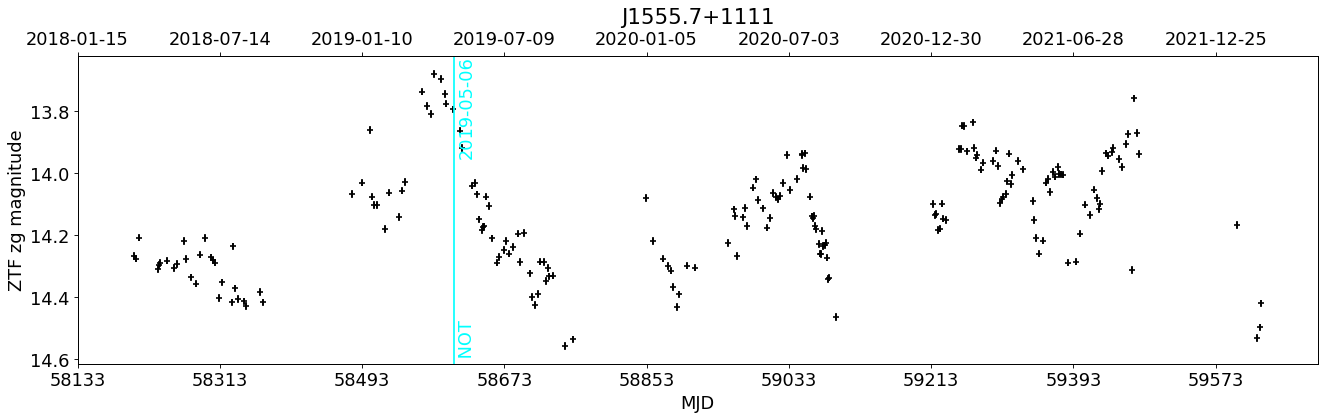}
   \includegraphics[width=0.9\textwidth]{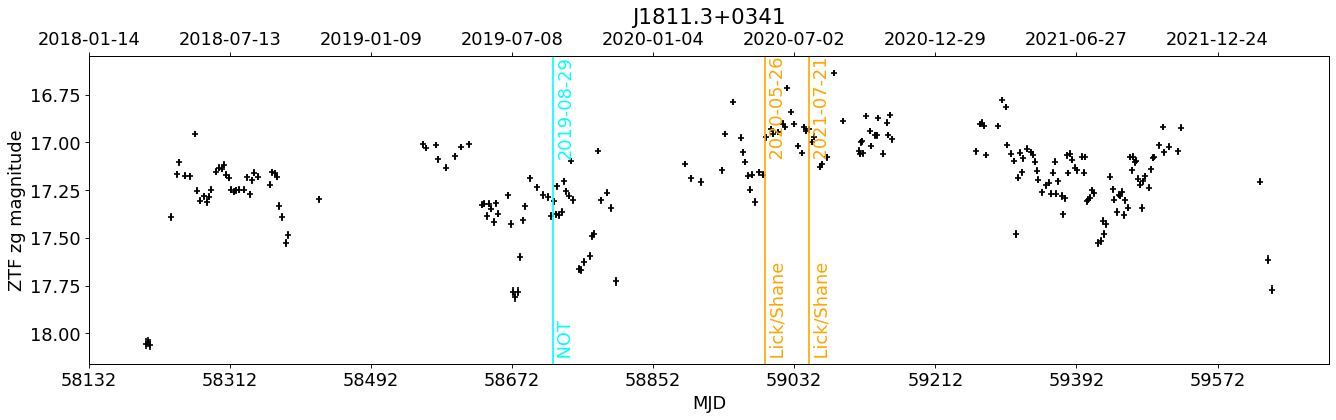}
   \caption{ZTF g-band light curves of  J1150.5+4154(top),  J1546.1+0818, J1555.7+1111 and J1811.3+0341 (bottom). The vertical lines show the timing of spectroscopy (orange) and our deep imaging observations (magenta).}
              \label{LC2}
    \end{figure*}

   \begin{figure*}
   \centering
   \includegraphics[width=0.9\textwidth]{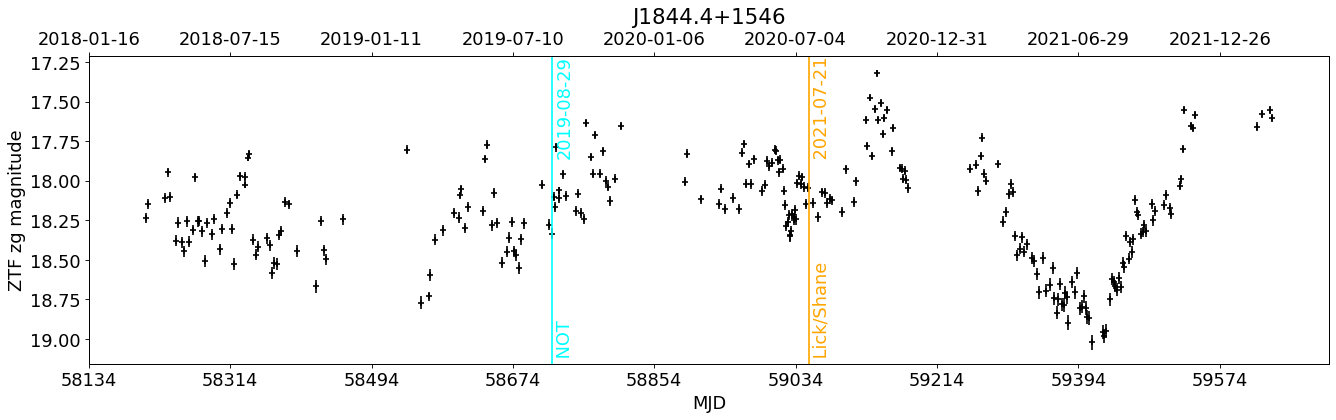}
   \includegraphics[width=0.9\textwidth]{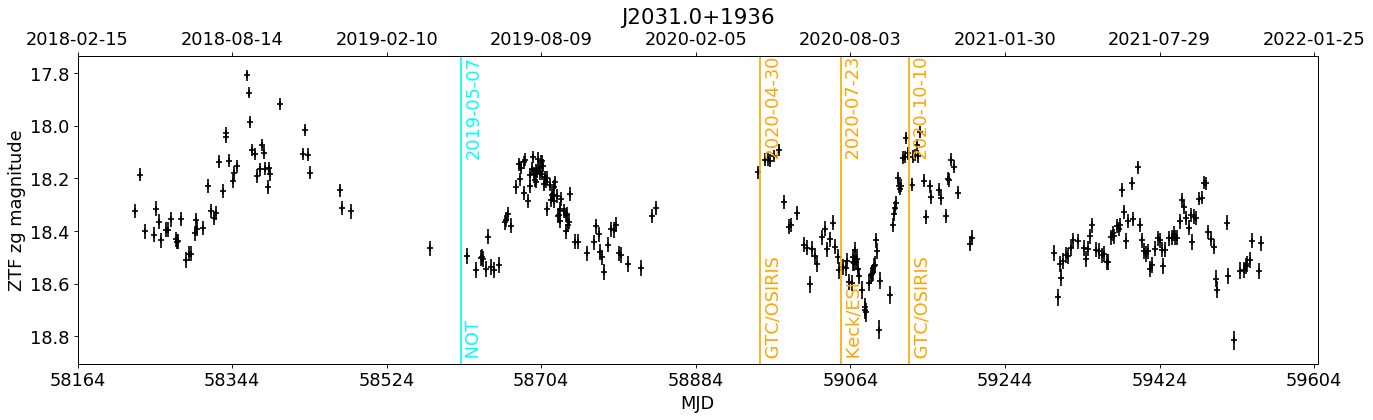}
   \includegraphics[width=0.9\textwidth]{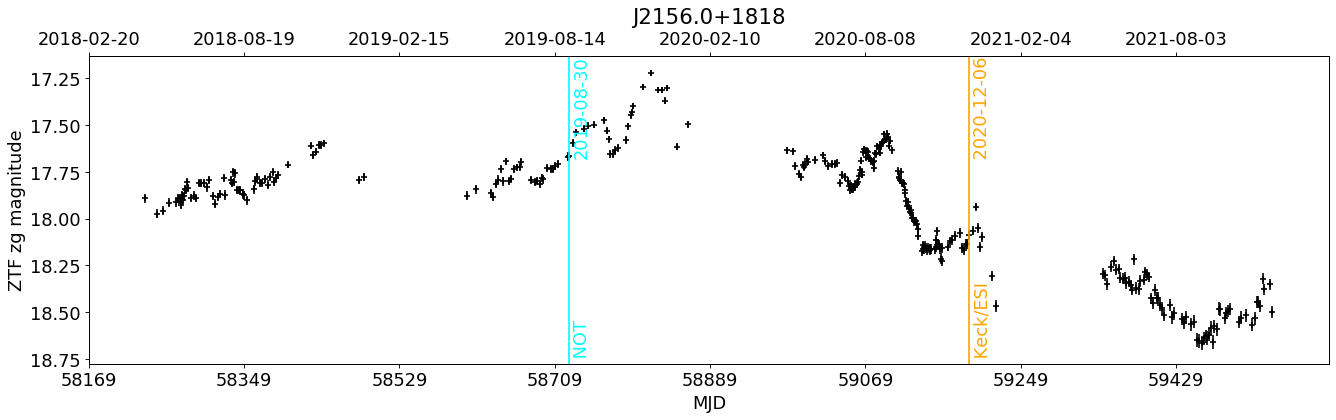}
   \includegraphics[width=0.9\textwidth]{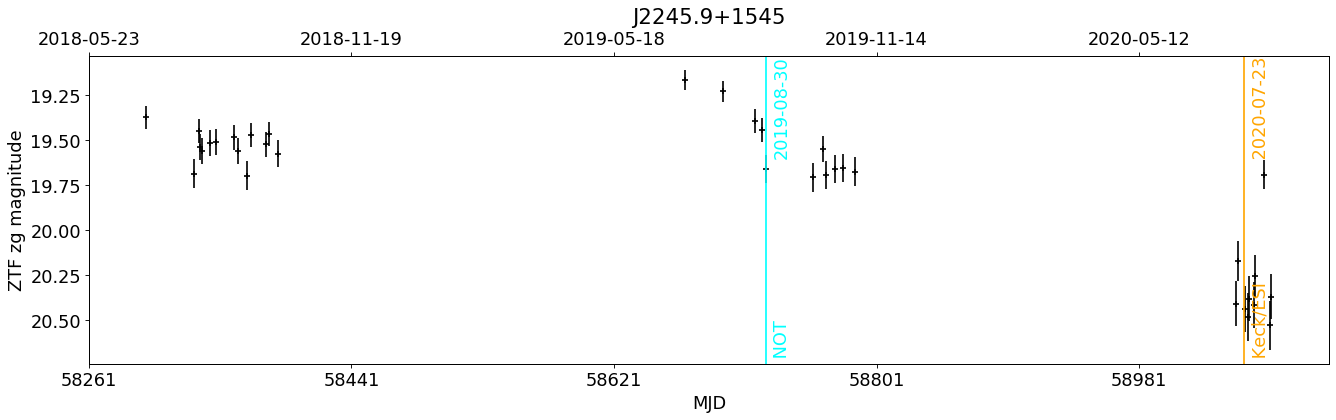}
    \caption{ZTF g-band light curves of J1844.4+1546(top),  J2031.0+1936, J2156.0+1818 and J2245.9+1545(bottom). The vertical lines show the timing of spectroscopy (orange) and our deep imaging observations (magenta).}
              \label{LC3}
              \end{figure*}

\section{Companion galaxies from SDSS}
The redshifts of the galaxies within 7.5 arcmin of the target blazar for all blazars of our sample that are within SDSS footprint~\ref{tab_companions_1},\ref{tab_companions_2}, \ref{tab_companions_3} and \ref{tab_companions_4}. We selected this radius, because at z=0.3 this radius corresponds to approximately 2\,Mpc, which is a typical search radius for companion galaxies.
    
\begin{table*}
\caption{\label{tab_companions_1}Nearby galaxies with spectroscopic redshift from SDSS}          
\centering  
\begin{tabular}{lccccr}
\hline 
\multicolumn{1}{c}{(1)} & (2) & (3) & (4) & (5) & (6)\\
\multicolumn{1}{c}{{3FHL name}}  & z$_{source}$ & RA & dec & z$_{galaxy}$ & separation\\  
\hline\\
J0045.3+2127 & 0.4253 & & & \\
&&00:45:18.87& +21:27:48.08& 0.426467&    10.0\\
&&00:45:20.02& +21:26:45.52& 0.421512&    55.4\\
&&00:45:15.84& +21:26:51.28& 0.425774&    68.5\\
&&00:45:11.94& +21:28:11.57& 0.620532&   107.5\\
&&00:45:22.60& +21:26:02.00& 0.321634&   108.3\\
&&00:45:20.47& +21:29:51.39& 0.418936&   132.4\\
&&00:45:28.30& +21:28:20.41& 0.062705&   132.0\\
&&00:45:12.44& +21:26:03.95& 0.430495&   135.7\\
&&00:45:08.12& +21:28:23.96& 0.400371&   162.1\\
&&00:45:17.23& +21:24:51.63& 0.321582&   170.8\\
&&00:45:20.01& +21:31:46.68& 0.709595&   246.9\\
&&00:45:00.22& +21:24:33.11& 0.482452&   325.4\\
\hline
J0905.5+1357 & 0.2239 or 0.644 & & & \\
&&09:05:35.057 &+13:58:20.107 &0.6385 &13.8\\
&&09:05:30.864 &+14:00:07.402 &0.1641 &135.1\\
&&09:05:41.167 &+13:56:04.488 &0.4503 &151.5\\
&&09:05:48.715 &+13:58:47.626 &0.3727 &204.0\\
&&09:05:19.930 &+13:58:28.542 &0.7636 &220.3\\
&&09:05:51.598 &+13:57:11.437 &0.3724 &247.9\\
&&09:05:26.714 &+13:54:28.249 &0.3107 &249.1\\
&&09:05:50.856 &+13:59:47.058 &0.3712 &252.0\\
&&09:05:46.697 &+14:01:59.858 &0.1784 &289.1\\
&&09:05:32.832 &+13:53:06.436 &0.6416 &301.5\\
&&09:05:56.551 &+13:56:43.235 &0.5483 &324.71\\
&&09:05:44.885 &+13:53:06.850 &0.3909 &332.34\\
&&09:05:12.487 &+13:59:54.107 &0.6935 &344.78\\
&&09:05:46.690 &+14:03:22.147 &0.8126 &358.80\\
&&09:05:29.191 &+14:04:35.119 &0.3123 &397.83\\
&&09:05:39.019 &+13:51:24.998 &0.3900 &405.60\\
&&09:05:52.054 &+14:03:37.004 &0.3255 &413.56\\
&&09:05:53.275 &+13:52:42.002 &0.0685 &419.61\\
\hline
\end{tabular}
\tablefoot{Columns: (1) source name in 3FHL catalogue (2) spectroscopic redshift of the source (3) right ascension of the nearby galaxy (4) declination of the nearby galaxy  (5) redshift of the nearby galaxy. (6) separation between the blazar and galaxy in arcsec.}
\\
\end{table*}

\begin{table*}
\caption{\label{tab_companions_2}Nearby galaxies with spectroscopic redshift from SDSS (continued)}          
\centering  
\begin{tabular}{lccccr}
\hline 
\multicolumn{1}{c}{(1)} & (2) & (3) & (4) & (5) & (6)\\
\multicolumn{1}{c}{{3FHL name}}  & z$_{source}$ & RA & dec & z$_{galaxy}$ & separation\\  
\hline\\
J0915.9+2933 & & & & \\
&&09:15:52.553 &+29:32:51.058 &0.4886 &33.0\\
&&09:15:56.818 &+29:33:9.691 &0.5280 &59.4\\
&&09:15:56.990 &+29:33:47.380 &0.4871 &64.3\\
&&09:15:53.614 &+29:32:9.856 &0.5308 &75.9\\
&&09:15:46.020 &+29:32:0.647 &0.5942 &117.9\\
&&09:15:51.226 &+29:31:8.731 &0.1809 &136.2\\
&&09:16:02.798 &+29:33:11.736 &0.5302 &136.2\\
&&09:15:50.453 &+29:35:39.426 &0.1857 &137.7\\
&&09:15:55.171 &+29:31:9.232 &0.4243 &139.6\\
&&09:16:03.487 &+29:33:12.445 &0.5321 &145.1\\
&&09:15:59.945 &+29:35:30.221 &0.7452 &160.0\\
&&09:15:52.850 &+29:36:18.529 &0.5308 &174.6\\
&&09:15:52.303 &+29:36:38.419 &0.5322 &194.4\\
&&09:15:47.851 &+29:36:49.381 &0.5316 &213.7\\
&&09:15:34.733 &+29:32:29.987 &0.9701 &236.8\\
&&09:16:03.228 &+29:29:51.907 &0.5323 &254.9\\
&&09:15:59.791 &+29:37:32.707 &0.5310 &266.7\\
&&09:15:52.642 &+29:28:54.048 &0.7465 &270.0\\
&&09:16:08.446 &+29:36:40.428 &0.1226 &287.0\\
&&09:15:33.624 &+29:36:6.710 &0.2207 &294.0\\
&&09:16:06.506 &+29:37:15.168 &0.8410 &295.4\\
&&09:16:15.456 &+29:32:23.852 &0.5984 &306.8\\
&&09:15:57.502 &+29:28:22.415 &0.5266 &308.9\\
&&09:15:32.918 &+29:36:28.645 &0.7966 &314.1\\
&&09:15:42.578 &+29:38:32.377 &0.5332 &333.9\\
&&09:15:52.212 &+29:39:1.134 &0.1027 &337.1\\
&&09:16:00.634 &+29:28:4.505 &0.5255 &337.1\\
&&09:15:43.574 &+29:27:55.858 &0.9368 &347.8\\
&&09:16:01.063 &+29:27:50.094 &0.5306 &352.6\\
&&09:15:44.945 &+29:39:23.922 &0.1855 &372.8\\
&&09:15:28.356 &+29:37:1.740 &0.7279 &381.8\\
&&09:15:23.623 &+29:35:0.960 &0.3016 &387.8\\
&&09:16:07.735 &+29:27:42.023 &0.5327 &396.3\\
&&09:15:54.151 &+29:26:45.715 &0.4924 &399.0\\
&&09:16:06.017 &+29:27:13.291 &0.4983 &411.2\\
&&09:16:24.396 &+29:33:28.015 &0.4239 &417.5\\
&&09:15:33.175 &+29:27:43.859 &0.1862 &422.7\\
&&09:15:57.782 &+29:26:26.459 &0.1220 &423.5\\
&&09:15:44.366 &+29:40:41.045 &0.6324 &449.4\\
\hline
\end{tabular}
\tablefoot{Columns: (1) source name in 3FHL catalogue (2) spectroscopic redshift of the source (3) right ascension of the nearby galaxy (4) declination of the nearby galaxy  (5) redshift of the nearby galaxy. (6) separation between the blazar and galaxy in arcsec.}
\\
\end{table*}

\begin{table*}
\caption{\label{tab_companions_3}Nearby galaxies with spectroscopic redshift from SDSS (continued)}          
\centering  
\begin{tabular}{lccccr}
\hline 
\multicolumn{1}{c}{(1)} & (2) & (3) & (4) & (5) & (6)\\
\multicolumn{1}{c}{{3FHL name}}  & z$_{source}$ & RA & dec & z$_{galaxy}$ & separation\\  
\hline\\
J1150.5+4154 &$0.42\pm0.09$& & & \\
&&11:50:36.214 &+41:54:32.450 &0.3219 &18.0\\
&&11:50:34.855 &+41:55:07.255 &0.3268 &27.1\\
&&11:50:32.170 &+41:55:34.568 &0.3272 &61.6\\
&&11:50:27.955 &+41:55:22.228 &0.5970 &86.8\\
&&11:50:32.371 &+41:56:05.525 &0.3250 &89.4\\
&&11:50:42.305 &+41:55:33.946 &0.3217 &100.0\\
&&11:50:22.380 &+41:54:50.735 &0.3608 &138.5\\
&&11:50:25.661 &+41:52:57.367 &0.3271 &144.5\\
&&11:50:34.848 &+41:51:58.496 &0.3807 &161.6\\
&&11:50:25.867 &+41:56:49.470 &0.3252 &163.0\\
&&11:50:20.578 &+41:53:58.740 &0.3652 &163.6\\
&&11:50:31.829 &+41:57:24.732 &0.3253 &167.8\\
&&11:50:48.744 &+41:53:04.834 &0.3266 &183.0\\
&&11:50:18.442 &+41:54:01.580 &0.3621 &186.1\\
&&11:50:47.858 &+41:56:58.250 &0.3267 &201.2\\
&&11:50:30.473 &+41:58:00.800 &0.3288 &206.3\\
&&11:50:53.210 &+41:55:35.159 &0.7656 &213.2\\
&&11:50:15.578 &+41:54:26.496 &0.3818 &214.5\\
&&11:50:17.412 &+41:56:19.828 &0.3625 &217.7\\
&&11:50:44.470 &+41:58:08.357 &0.3557 &234.8\\
&&11:50:45.125 &+41:58:04.649 &0.6549 &235.0\\
&&11:50:13.882 &+41:53:01.162 &0.0881 &253.2\\
&&11:50:46.934 &+41:50:54.812 &0.8034 &263.2\\
&&11:50:11.328 &+41:54:02.070 &0.5988 &264.3\\
&&11:50:15.847 &+41:50:42.788 &0.9279 &317.7\\
&&11:50:34.716 &+41:49:20.528 &0.4279 &319.6\\
&&11:50:07.111 &+41:53:07.721 &0.3900 &322.2\\
&&11:50:16.092 &+41:59:13.859 &0.3821 &343.9\\
&&11:51:04.462 &+41:52:58.508 &0.7068 &346.9\\
&&11:50:03.782 &+41:53:58.661 &0.4636 &348.2\\
&&11:50:07.536 &+41:58:10.909 &0.3627 &369.7\\
&&11:49:58.819 &+41:57:26.309 &0.0822 &434.0\\
\hline
J1546.1+0818 & & & & \\
&&15:46:04.963 &+08:18:28.426 &0.5130 &46.2\\
&&15:46:09.413 &+08:19:11.418 &0.0703 &76.6\\
&&15:45:54.648 &+08:18:29.725 &0.5516 &149.1\\
&&15:45:54.101 &+08:18:38.553 &0.5617 &154.6\\
&&15:46:12.605 &+08:22:41.919 &0.0407 &242.6\\
&&15:46:17.738 &+08:16:23.605 &0.0701 &262.5\\
&&15:45:56.302 &+08:14:59.339 &0.0418 &280.1\\
&&15:46:03.281 &+08:14:26.052 &0.1241 &287.7\\
&&15:46:19.301 &+08:22:20.667 &0.4866 &291.5\\
&&15:46:23.918 &+08:19:54.791 &0.5076 &294.8\\
&&15:46:07.922 &+08:14:22.628 &0.5717 &295.8\\
&&15:46:07.572 &+08:24:06.090 &0.3548 &296.8\\
&&15:45:44.873 &+08:17:21.636 &0.2020 &308.6\\
&&15:45:43.118 &+08:15:53.668 &0.0707 &371.9\\
&&15:45:46.430 &+08:23:37.036 &0.6908 &373.4\\
&&15:46:27.041 &+08:15:29.236 &0.0677 &405.8\\
\hline
\end{tabular}
\tablefoot{Columns: (1) source name in 3FHL catalogue (2) imaging redshift of the source (3) right ascension of the nearby galaxy (4) declination of the nearby galaxy  (5) redshift of the nearby galaxy. (6) separation between the blazar and galaxy in arcsec.}
\\
\end{table*}

\begin{table*}
\caption{\label{tab_companions_4}Nearby galaxies with spectroscopic redshift from SDSS (continued)}          
\centering  
\begin{tabular}{lccccr}
\hline 
\multicolumn{1}{c}{(1)} & (2) & (3) & (4) & (5) & (6)\\
\multicolumn{1}{c}{{3FHL name}}  & z$_{source}$ & RA & dec & z$_{galaxy}$ & separation\\  
\hline\\
J1555.7+1111 & & & & \\
&&15:55:43.951 &+11:11:56.796 &0.4343 &35.1\\
&&15:55:33.238 &+11:10:11.176 &0.5054 &161.8\\
&&15:55:58.022 &+11:11:44.030 &0.0415 &221.3\\
&&15:55:42.890 &+11:7:40.019 &0.1520 &224.4\\
&&15:55:26.585 &+11:11:54.992 &0.1056 &244.1\\
&&15:55:53.095 &+11:7:23.851 &0.4320 &282.4\\
&&15:55:59.237 &+11:8:48.437 &0.1513 &284.8\\
&&15:56:0.581 &+11:13:57.014 &0.1354 &299.8\\
&&15:55:21.055 &+11:8:39.419 &0.0450 &363.2\\
&&15:56:1.368 &+11:15:52.780 &0.1344 &380.4\\
&&15:55:18.355 &+11:14:49.841 &0.1516 &417.3\\
&&15:55:19.390 &+11:7:26.260 &0.0707 &421.8\\
\hline
J2156.0+1818 & $0.60\pm0.10$(i) & & & \\
&&21:56:08.570 &+18:19:53.508 &1.0803 &124.8\\
&&21:56:12.041 &+18:17:13.772 &0.6334 &170.0\\
&&21:56:15.358 &+18:20:25.573 &0.1837 &223.4\\
&&21:55:53.494 &+18:22:42.362 &0.3181 &271.3\\
&&21:55:44.544 &+18:16:9.005 &0.4826 &285.0\\
&&21:55:59.993 &+18:13:44.101 &0.3082 &294.0\\
&&21:56:23.405 &+18:16:12.281 &0.6711 &342.1\\
&&21:55:53.318 &+18:24:13.424 &0.5136 &356.5\\
&&21:56:28.831 &+18:19:18.649 &0.5336 &389.4\\
&&21:55:51.967 &+18:11:42.140 &0.5123 &437.3\\
&&21:56:17.321 &+18:24:57.895 &0.5107 &441.4\\
\hline
J2245.9+1545 &0.5965(s) & & & \\
&&22:45:55.356 &+15:45:28.487 &0.4347 &148.8\\
&&22:46:07.939 &+15:41:52.217 &0.3241 &168.6\\
&&22:46:00.634 &+15:48:04.010 &0.4939 &217.9\\
&&22:45:58.769 &+15:48:29.354 &0.6065 &250.6\\
&&22:45:46.406 &+15:43:46.387 &0.4334 &272.7\\
&&22:45:44.330 &+15:45:32.278 &0.2035 &303.6\\
&&22:45:48.482 &+15:47:44.052 &0.6414 &303.9\\
&&22:45:45.475 &+15:46:35.065 &0.5197 &306.1\\
&&22:45:45.780 &+15:48:09.752 &0.3238 &350.5\\
&&22:45:55.370 &+15:50:12.350 &0.4318 &364.4\\
&&22:45:37.039 &+15:45:25.632 &0.2334 &406.6\\
&&22:46:31.896 &+15:46:43.226 &0.4326 &408.9\\
&&22:45:45.797 &+15:49:51.625 &0.6259 &420.4\\
&&22:46:33.732 &+15:46:44.278 &0.4581 &434.5\\
\hline
\end{tabular}
\tablefoot{Columns: (1) source name in 3FHL catalogue (2) redshift of the source, (s)=spectroscopic, (i)=imaging (3) right ascension of the nearby galaxy (4) declination of the nearby galaxy  (5) redshift of the nearby galaxy. (6) separation between the blazar and galaxy in arcsec.}
\\
\end{table*}

\end{appendix}
\end{document}